\documentstyle[aps,pre]{revtex}
\begin{document}
\draft

\title{Magnetoplasmon Excitations and Spin Density Instabilities
in an Integer Quantum Hall System with a Tilted Magnetic Field}

\author{Daw-Wei Wang$^1$, S. Das Sarma$^1$, Eugene Demler$^2$, 
Bertrand I. Halperin$^2$}

\address{$^1$Condensed Matter Theory Group, 
Department of Physics, University of Maryland,
College Park, MD 20770\\
$^2$Physics Department, Harvard University, Cambridge, MA 02138}

\date{\today}

\maketitle


\begin{abstract}

We study the magnetoplasmon collective mode excitations of
integer quantum Hall systems in a parabolically confined quantum well
nanostructure in the presence of a tilted magnetic field
by using the time-dependent Hartree-Fock approximation.
For even integer filling, we find
that the dispersion of a spin density mode has a magneto-roton minimum
at finite wavevectors, at a few times $10^6$ cm$^{-1}$ for parallel
fields of order 1-10 Tesla, {\it only} in the direction
perpendicular to the in-plane magnetic field, while
the mode energy increases monotonously with wavevector parallel to
the in-plane magnetic field. 
When the in-plane magnetic field is strong enough (well above 10 
Tesla),we speculate that this roton minimum
may reach zero energy, suggesting
a possible second order phase transition to a state with broken 
translational and spin symmetries. We discuss the possibility for observing
such parallel field-induced quantum phase transitions. We also
derive an expression for the dielectric function within the
time-dependent Hartree-Fock approximation and include screening
effects in our magnetoplasmon calculation. 
We discuss several exotic symmetry-broken phases that may be stable
in finite parallel fields, and propose
that the transport anisotropy, observed recently
in parallel field experiments,
may be due to the formation of a skyrmion stripe phase
predicted in our theory.
Our predicted anisotropic finite
wavevector suppression, perhaps even a mode-softening leading to the quantum
phase transition to the anisotropic phase, in the collective spin excitation
mode of the wide well system in the direction transverse to the applied
parallel magnetic field should be directly experimentally observable via the
inelastic light scattering spectroscopy.
\end{abstract}

\pacs{PACS numbers: 73.43.-f, 73.43.Nq, 73.43.Lp, 73.43.Cd}
\vfill\eject
\vskip 1pc
\section{Introduction}
\label{introduction}

Observations of integral quantum Hall effect (IQHE, 1980) and 
fractional quantum Hall effect (FQHE, 1982)  
are important landmarks of condensed matter physics in recent
decades \cite{review}. In quantum Hall systems, electrons are "frozen"
(in their orbital motion)
in discrete Landau levels by the external magnetic field, and 
have gapped excitations at integer
or fractional filling factors.
There is considerable richness of the
phase diagram when additional (i.e. in addition to the orbital motion)
degrees of freedom associated with spin, layer,
or subband index are introduced \cite{review,jungwirth01,girvin}.
These multicomponent quantum Hall systems have been extensively studied
both theoretically and experimentally in recent years. In general,
since the spin (Zeeman) energy is much smaller
than the cyclotron energy due to the small effective $g$-factor and the small
effective mass of electrons in GaAs based QH systems, 
the spin degree of freedom is not important energetically
compared to the orbital motion. 
But spin can be crucial when a second quantum Hall system is 
coupled coherently (for example, in a double quantum well [DQW] system) or 
an additional magnetic field is applied in the direction parallel to the
two-dimensional (2D) semiconductor quantum well plane.
In the first situation, the
finite barrier energy between the two wells opens a 
gap ($\Delta_{SAS}$) between
a symmetric and an antisymmetric subbands, which can be tuned
by electron tunneling, layer separation, and/or bias 
voltage \cite{jungwirth01}.
When $\Delta_{SAS}$ is close to the Zeeman splitting energy,
interesting physics has been predicted theoretically 
\cite{canted_phase} and 
observed experimentally \cite{caf_exp1,caf_exp2}.
On the other hand, physics of the second situation, where a tilted
magnetic field is applied to a wide width well (WWW) system to couple
subbands of a wide well with spin-split Landau levels, has not yet
been extensively explored. One reason for this
is that the strength of the applied tilted magnetic field has to be
very large ($> 25$ Tesla) in order to sufficiently enhance the Zeeman energy
to be comparable to the Landau level separation in GaAs. Such strong and
uniform magnetic fields has only been available  
very recently \cite{pan01}. From a theoretical point of view,
studying QH effects in a WWW with tilted magnetic field is
difficult because 
the in-plane magnetic field hybridizes the 2D electron subbands
arising from the
confinement potential in the growth direction (i.e. perpendicular
to the 2D plane) with the orbital Landau 
levels so that
the electron wavefunction of a WWW is a complicated combination of
electric (``subbands'') and magnetic (``Landau levels'') quantization
even at the single particle level.
It is sometimes simplistically believed that, if parameters are chosen properly
in an isospin language, then
a WWW system in a tilted field (at least for) the closest two Landau
levels near the degeneracy point could be approximately mapped onto a
DQW system. We emphasize that this mapping is
not exact and misses subtle and interesting physics associated with a WWW
in a tilted field.
For example, experimentally a WWW in a tilted field is found to display
both three-dimensional (3D) and two-dimensional properties \cite{sergio01}.
In some situations a WWW system could behave very much like a DQW system
albeit with strong tunneling) \cite{suen91_ww_to_dw}.
More strikingly, the recent observation of anisotropic resistance
at even filling factors in a WWW system with an in-plane field
\cite{pan01} shows a possible stripe phase formation induced by
electron-electron interaction near a degeneracy or a level crossing point.

Inspired by the observed anisotropic transport properties at integer filling
factors \cite{pan01}, we   
investigate in this paper the collective mode excitations of 
integer quantum Hall systems in a wide quantum well with a tilted 
magnetic field (i.e. in the presence of an in-plane magnetic field)
by using the time-dependent Hartree-Fock approximation (TDHFA).
We extend the work of Kallin and Halperin \cite{kallin}
for a strictly 2D system, i.e. a zero width well (ZWW) to a WWW system
and derive a full analytical expression for the mode dispersion energy.
To keep our theory analytically tractable we choose our quantum well 
confinement potential to be parabolic ("parabolic well"). 
Our choice of parabolic confinement is dictated by the fact 
that the corresponding single-particle problem (i.e. an electron moving
in a one-dimensional parabolic potential along the $z$-direction in the 
presence of an arbitrary magnetic field) can be exactly analytically
solved enabling essentially a complete analytic solution of the many-body
TDHF solution of the collective mode spectra (essentially on the same 
footing as the 2D Kallin-Halperin work in Ref. \cite{kallin}) in the
WWW system in the presence of a tilted magnetic field (i.e. both
the in-plane field and the perpendicular field producing the Landau 
quantization). The work presented in this paper is therefore a direct
(and highly non-trivial) generalization of the strictly 2D Kallin-Halperin 
work \cite{kallin} on the magnetoplasmons of a 2D electron gas (in the presence
of only a perpendicular magnetic field) to a parabolic
WWW system in the presence of a tilted magnetic field.
We study both charge and spin mode collective excitations in systems of
different electron densities, magnetic field strengths, and well widths.
At even integer factors, we find that
the dispersion of spin density mode has a magneto-roton minimum
{\it only} at a finite wavevector in the direction
perpendicular to the in-plane magnetic field, while
it increases monotonously with respect to the wavevector parallel to
the in-plane magnetic field.
When the in-plane magnetic field is sufficiently strong, this roton minimum
may reach zero energy before the ground state becomes polarized, suggesting
a possible second order phase transition to a state with broken
translational and spin symmetries.
The possibility of this quantum phase transition (to an anisotropic
symmetry-broken state) in the presence of a tilted field is one main 
new result of our work. We also
derive the full formula for the dielectric function of the system within
TDHFA by including the ladder diagrams consistently, so that it can be applied
to other systems even when only few Landau levels are occupied. 
We include such screening in our collective mode calculation and discuss
its effect to the magneto-roton minimum.  

Before jumping into the details of the collective mode calculation,
it is instructive to discuss in the appropriate context some earlier work 
in parabolic wells and in the ground state instability (i.e.
the softening of collective modes) of similar systems.
Among the models of finite width wells, parabolic wells
are considered special, because the electron gas,
in screening the parabolic conduction band edge potential, forms a constant
density slab, being a good approximation to a 3D 
jellium where electrons move in
a constant positive background charge density \cite{wide_well}. 
Furthermore, the parabolic
confinement potential can be exactly diagonalized in a center-of-mass
coordinate and therefore gives a non-spin-flip optical absorption energy
exactly the same as its noninteracting result in the long wavelength
limit (the so called generalized Kohn's theorem) \cite{brey89,yip}. 
As mentioned above,we use a parabolic confinement 
potential, because
it allows us to find simple noninteracting eigenstates in the presence
of a tilted magnetic field, which then provides a good starting point
to consider many-body effects. The effects of {\it imperfect} parabolic 
confinement potential on the collective excitations
have earlier been studied either with only a perpendicular
magnetic field \cite{brey90_imprefect} or with only an in-plane magnetic field
\cite{tamborenea94_imprefect}. Only rather small quantitative
corrections were found (for example, small shift of resonance energy, and
slight broadening of the absorption peak)
for realistic wells (which necessarily deviate from ideal parabolic 
confinement considered in our work). We believe, therefore, that our 
theoretical results should apply with quantitative accuracy to realistic 
parabolic quantum wells, and qualitatively to rectangular quantum wells
\cite{imprefect_parabolic}.

It is generally believed
that in both three and two dimensions, when an 
infinitely strong magnetic field is applied,
electrons undergoes a phase transition to a Wigner crystal state with
broken translational symmetry at low temperatures.
In the intermediate magnetic
field region, Celli and Mermin \cite{celli65_3d} proposed
a long time ago a possible
exchange induced spin-density-wave (SDW) instability in a three-dimensional
electron system.
More recently, GaAs based semiconductor wide parabolic wells
have been proposed as good candidates for observing such SDW 
instabilities since wide parabolic wells are essentially ideal 3D 
electron systems \cite{sergio01,hembree94,brey89_sdw}.
Brey and Halperin \cite{brey89_sdw} proposed that 
the SDW instability and the transport anisotropy should 
be observed in a wide parabolic semiconductor quantum well system
when an intermediate in-plane magnetic field is applied.
Similarly, correlation-driven intersubband SDW instability has been predicted 
by Das Sarma and Tamborenea
in DQW systems at low carrier densities \cite{dassarma94_sdw_dqw}.
Intersubband-induced charge-density-wave (CDW) instability
in a wide parabolic well with a
perpendicular magnetic field was also investigated \cite{brey91_cdw}.
To the best of our knowledge, however, these 
theoretically proposed (translational symmetry
breaking) instabilities have not yet been observed experimentally.
The only two experimentally observed candidates for charge (or spin) 
density wave instability
in a quantum Hall system
are the stripe phases (and the associate liquid crystal phases
\cite{fogler01_liquid_crystal}) 
in high half-odd-integer quantum Hall systems
($\nu=9/2$, 11/2, etc)
\cite{stripe_HOI} with or without in-plane magnetic field, and the 
stripe phases observed in an integer quantum Hall system
in a wide well
subject to a strong tilted magnetic field \cite{pan01}. Although the
ground state of the former system has been extensively studied 
\cite{fogler01_liquid_crystal} and is
generally believed to be a "unidirectional
coherent charge density wave" \cite{jungwirth99,brey00}, 
the transport anisotropy 
in the wide well with a tilted magnetic field \cite{pan01} is not
yet understood and not much theoretical work has appeared on this problem
except for our recent short communication \cite{eugene01}. 
Our recent work \cite{eugene01} based on 
Hartree-Fock (HF) calculation in a DQW system shows that spin-charge-texture 
(skyrmion) stripe could be the possible ground state for a WWW
system, providing a possible explanation for the observed transport anisotropy
in Ref. \cite{pan01}. In this paper, for the first time we 
show the complete analytical
and numerical work in calculating the collective magnetoplasmon 
mode dispersion within TDHFA and the observed mode softening confirms
the existence of a novel phase proposed in Ref. \cite{eugene01}.

This paper is organized as follows. In Sec. \ref{sing_particle}
we obtain the single-electron eigenstates in a parabolic
confinement potential with a tilted magnetic field. We first discuss
the noninteracting result in Sec. \ref{noninteracting} and then
the interacting (HF) result in Sec. \ref{single_HF}. 
In Sec. \ref{level_crossing} we show
that at even filling factors the system undergoes a first order 
phase transition from an unpolarized ground state for in-plane magnetic 
field, $B_\|<B_\|^\ast$, where $B_\|^\ast$ is a critical in-plane field
strength, to a polarized ground state for $B_\|>B_\|^\ast$. 
Based on the unpolarized integral quantum Hall 
ground state, the full theory with numerical results for the
magnetoplasmon dispersion (within TDHFA) are given 
in Sec. \ref{magnetoplasmon}.  
In Sec. \ref{screening}
we derive the TDHF dynamical dielectric function
for an integer quantum Hall system in a parabolic well with tilted 
magnetic field and use the result to study the magnetoplasmon dispersion
in screened TDHFA.
Implications of our results 
are discussed in Sec. \ref{discussion} and finally
we summarize our work in Sec. \ref{summary}.

\section{Single Electron Eigenstates and Ground State Energy}
\label{sing_particle}
\subsection{Non-interacting System}
\label{noninteracting}

We consider a parabolic confinement potential in $\hat{z}$ direction,
$U_{p}(z)= \frac{1}{2}~
m^* \omega_0^2 z^2$, where $m^*$ is 
the electron effective mass and $\omega_0$ is the confinement energy. 
A coordinate system is chosen such that the
perpendicular magnetic field $B_\bot$ is in $\hat{z}$ direction and the
parallel magnetic field $B_\|$ in $\hat{x}$ direction, with
the 2D electron system being in the $x-y$ plane. When the vector
potential is chosen in a Landau gauge, $\vec{A}=(0,B_\bot x-B_\| z,0)$,
the noninteracting single electron Hamiltonian can be written as
(we set $\hbar=1$ throughout this paper)
\begin{eqnarray}
{\cal H}_0& =& \frac{1}{2m^*}\left(\vec{p}+\frac{e \vec{A}}{c}\right)^2+
U_{p}(z)-g\mu_BB_{tot}S_z\nonumber\\
&=&\frac{p_x^2}{2m^\ast}+\frac{1}{2m^\ast}\left(p_y+\frac{eB_\perp x}{c}
-\frac{eB_\| z}{c}\right)^2+\frac{p_z^2}{2m^*}
+\frac{1}{2}m^* \omega_0^2 z^2-g\mu_BB_{tot}S_z
\label{noninteracting_H}
\end{eqnarray}
where $\mu_B$ is the Bohr magneton, and $g\sim 0.44$ for GaAs.
$S_z$ is the $z$-component of the spin operator 
along the total magnetic field,  
whose magnitude is $B_{tot}=\sqrt{B^2_\bot+B^2_\|}$.
$p_y$ is a good quantum number in this gauge and 
can be replaced by a constant $k$ (the guiding center 
coordinate). The remaining terms can 
be expressed by a $2\times 2$ matrix 
\begin{eqnarray}
{\cal H}_0 =
\frac{1}{2m^\ast}\left(p_x^2+p_z^2\right)
+\frac{m^\ast}{2}
[x',z]\cdot\left[\begin{array}{cc}
                  \omega_\bot^2 & -\omega_\bot\omega_\| \\
                 -\omega_\bot\omega_\| & \omega_b^2
                 \end{array}\right]
\cdot\left[\begin{array}{c}
           x' \\
           z
           \end{array}\right]
-\omega_z S_z
\label{Hamiltonian_unrotated}
\end{eqnarray}
where $\omega_{\bot,\|}={eB_{\bot,\|}}/{m^*c}$,
$\omega_b=\sqrt{\omega_0^2+\omega_\|^2}$,  $
\omega_z= g\mu_BB_{tot} $,
and $x'=x+\frac{ck}{eB_\bot}$.
Hamiltonian of Eq. (\ref{Hamiltonian_unrotated}) can be diagonalized
by a canonical transformation,
$[x',z]^T=\hat{U}(\theta)\cdot [\bar{x},\bar{z}]^T$ and
$[{p}_x,{p}_z]^T=\hat{U}(\theta)\cdot [\bar{p}_x,\bar{p}_z]^T$, with
\begin{equation}
\hat{U}(\theta)=\left[\begin{array}{cc}
                  \cos\theta & \sin\theta \\
                  -\sin\theta & \cos\theta
                 \end{array}\right],
\end{equation}
and $\tan(2\theta)=-2\omega_\bot\omega_\|/(\omega_b^2-\omega_\bot^2)$.
The new Hamiltonian describes two decoupled one-dimensional (1D) simple
harmonic oscillators in new coordinates, $\bar{x}$ and $\bar{z}$:
\begin{eqnarray}
\bar{\cal H}_0=\frac{1}{2m^\ast}\left(\bar{p}_x^2+\bar{p}_z^2\right)
+\frac{m^\ast\omega_1}{2}\bar{x}^2+\frac{m^\ast\omega_2}{2}\bar{z}^2
-\omega_z S_z,
\label{Hamiltonian_rotated}
\end{eqnarray}
where 
\begin{eqnarray}
\omega^2_{1,2}&=&\frac{1}{2}\left[\left(\omega_b^2+\omega_\bot^2\right)\pm
\sqrt{\left(\omega_b^2-\omega_\bot^2\right)^2+4\omega_\bot^2\omega_\|^2}
\,\right].
\label{omega12}
\end{eqnarray}
Using $(\vec{n},k,s)$ as eigenstate quantum numbers, where 
$\vec{n}=(n_1,n_2)$ is the orbital Landau level index and 
$s=\pm 1/2$ is the eigenvalues
of $S_z$, one obtains the noninteracting eigenenergies, $E^0_{\vec{n},s}$,
and eigenfunctions, $\phi^{0}_{\vec{n},k,s}(\vec{r})$:
\begin{equation}
E^0_{\vec{n},s}=\omega_1\left(n_1+\frac{1}{2}\right)+
\omega_2\left(n_2+\frac{1}{2}\right)-\omega_z s,
\label{energy_levels_eqn}
\end{equation}
and
\begin{eqnarray}
\phi^{0}_{\vec{n},k,s}(\vec{r})&=&\frac{e^{iky}}{\sqrt{L_y}}
\,\psi^{(1)}_{n_1}(\bar{x})\cdot
\psi^{(2)}_{n_2}(\bar{z})\nonumber\\
&=&\frac{e^{iky}}{\sqrt{L_y}}
\underbrace{{\psi}^{(1)}_{n_1}(\cos\theta(x+l_0^2k)-\sin\theta z)
\cdot{\psi}^{(2)}_{n_2}(\sin\theta(x+l_0^2k)+\cos\theta z)}_
{\displaystyle \equiv\Phi^0_{\vec{n},s}(x+l_0^2k,z)},
\label{noninteracting-wavefunctions}
\end{eqnarray}
where $L_y$ is the system length in $y$ direction and 
the function $\Phi^0_{\vec{n},s}(x+l_0^2k,z)$ has $x$ and $z$ components only.
$l_0\equiv\sqrt{1/m^*\omega_\bot}=\sqrt{c/eB_\bot}$ is the conventional
cyclotron radius.
We keep the spin index in $\Phi^0_{\vec{n},s}$ 
because these notations will later 
be generalized to an interacting system, where explicit spin dependence
may become crucial. In Eq. (\ref{noninteracting-wavefunctions}),
the function ${\psi}^{(i)}_{n}(x)$ is defined to be
\begin{eqnarray}
{\psi}^{(i)}_{n}(x)&=&\frac{1}{\sqrt{\pi^{1/2}2^nn!l_i}}
\exp\left[-\frac{x^2}{2l_i^2}\right]
H_n\left(\frac{x}{l_i}\right),  
\label{wf_i}
\end{eqnarray}
with $ l_i \equiv \sqrt{1/m^*\omega_i}$ for $ i=1,2$, and
$l_0\equiv\sqrt{1/m^*\omega_\bot}=\sqrt{c/eB_\bot}$ is the conventional
cyclotron radius. $H_n(x)$ is Hermite polynomial. 
It is instructive to consider the asymptotic form of the
eigenstate energies, Eq. (\ref{omega12}), and wavefunctions,
Eq. (\ref{noninteracting-wavefunctions}), in the 
following four extreme limits: (i)
Taking an infinite well width limit, $\omega_0\rightarrow 0$,
$\omega_1\rightarrow\sqrt{\omega_\|^2+\omega_\perp^2}$
and $\omega_2\rightarrow 0$ from Eq. (\ref{omega12}),
Eq. (\ref{Hamiltonian_rotated}) then shows that the free moving 
direction is restored along the ${\bar{z}}$ direction, which is 
perpendicular to the total magnetic field, $\vec{B}_{tot}$, showing
a 3D property. (ii) 
Taking a zero width limit 
($\omega_0\rightarrow\infty$), we have $\theta\rightarrow\pi/2$,
$\omega_1\rightarrow\omega_0\rightarrow\infty$, 
$\omega_2\rightarrow\omega_\perp$, and therefore
${\psi}^{(1)}_{n_1}(\bar{x})\rightarrow\sqrt{\delta(z)}$ and
${\psi}^{(2)}_{n_2}(\bar{x})\rightarrow{\psi}^{(0)}_{n_2}(x)$, the
usual orbital wavefunction of a 1D
simple harmonic oscillator.
Therefore by changing the value of $\omega_0$, one can obtain a quasi-2D
system, which has both pure 2D and 3D properties by taking different limits
of the confinement potential strength.
(iii) Similarly, for zero in-plane 
magnetic field limit ($\omega_\|\rightarrow 0$),
we have $\omega_1\rightarrow Max(\omega_\perp,\omega_0)$ and
$\omega_2\rightarrow Min(\omega_\perp,\omega_0)$, so that the
orbital motions in $x$ and $z$ direction are totally decoupled.
This is the usual (i.e. without an in-plane field) 
quantum Hall system in a parabolic well, whose 
collective mode dispersion has been studied in the literature 
\cite{brey91_cdw}. 
(iv) Finally we can take the strong parallel (in-plane) magnetic field limit
($B_\|\rightarrow\infty$), which is of interest in this paper.
In this limit, we have $\theta\rightarrow\pi/2$,
$\omega_1\rightarrow\omega_\|\rightarrow\infty$, and $\omega_2\rightarrow 
\omega_0\omega_\perp/\omega_\|\to 0$, i.e.  
the in-plane magnetic field 
enhances the effective confinement of a wide well system 
(compared to (ii)) and therefore a
WWW system with a strong parallel field becomes similar to a
thin well (strictly 2D) system with small Landau level 
energy separation.
We emphasize, however, that our results shown 
below apply for any finite strength of $B_\|$ valid to
the lowest order of the ratio of the interaction strength to
the noninteracting level separation. We will consider the strong
in-plane magnetic field limit only when studying the screening effect 
in Sec. \ref{screening}.

Energy levels described by Eq. (\ref{energy_levels_eqn})
are shown in Fig. \ref{energy_levels_figure}
as a function of
in-plane magnetic field for a choice of parameters 
similar to the experimental samples in \cite{pan01}:
electron density $n_e=0.42\times 10^{12}$ cm$^{-2}$, $m^*=0.07~m_0$ 
($m_0$ is the bare electron mass) and $\omega_0 =
7$ meV. The confinement energy is such that the size
of the first subband electron wavefunction in zero field is 260 \AA. 
The perpendicular magnetic field, $B_\perp$, 
is chosen to be 2.97 Tesla for $\nu=2$.

Using the noninteracting single particle wavefunction 
in Eq. (\ref{noninteracting-wavefunctions}),
the noninteracting single electron Green's function can be easily obtained:
\begin{eqnarray}
&&G^0(\vec{r}_1,\tau_1;\vec{r}_2,\tau_2)=\sum_{\sigma}\sum_{\vec{n}_\sigma}
G^0_{\vec{n},\sigma}(\vec{r}_1,\tau_1;\vec{r}_2,\tau_2)  \nonumber\\
&=&\sum_{\sigma}\sum_{\vec{n}_\sigma}
\sum_k{\phi^0_{\vec{n},k,\sigma}}^\dagger(\vec{r}_2)
\phi^0_{\vec{n},k,\sigma}(\vec{r}_1)
\underbrace{
e^{(\tau_2-\tau_1)(E^0_{\vec{n},\sigma}-\mu)}
\left[\theta(\tau_2-\tau_1)\theta(\mu
-E^0_{\vec{n},\sigma})
-\theta(\tau_1-\tau_2)\theta(E^0_{\vec{n},\sigma}-
\mu)\right]}
_{\displaystyle \equiv{\cal G}_{\vec{n},\sigma}^{0}(\tau_1-\tau_2)},
\label{Greenfunc2}
\end{eqnarray}
where $\tau_i$ is the imaginary time, and 
$\mu$ is chemical potential at zero temperature. 
The Heaviside theta function $\theta(x)=1$ for $x\geq 0$ and is zero otherwise.

\subsection{Interacting System in Hartree-Fock Approximation}
\label{single_HF}

When electron-electron interaction is considered, we use self-consistent
Hartree-Fock approximation (SCHFA) to calculate
the single electron wavefunction self-consistently by including the Hartree and
Fock potentials in the single particle Hamiltonian. This approximation
is the standard leading-order many-body (self-consistent) expansion in the
(unscreened) Coulomb interaction \cite{fetter} whose one-loop 
Feynman diagram
representation is shown in Fig. \ref{feynm_figure}(a).
In SCHFA, the wave 
equation for the quantum Hall system is \cite{kittel}:
\begin{eqnarray}
{E}_{\vec{n},k,\sigma}{\phi}_{\vec{n},k,\sigma}(\vec{r}\,)
&=&\left[{\cal H}_0+\int d\vec{r}\,'V(\vec{r}-\vec{r}\,')
\sum_{\vec{m},p,s}\nu_{m,p,s}
{\phi}_{\vec{m},p,s}{}^\dagger(\vec{r}\,')
{\phi}_{\vec{m},p,s}(\vec{r}\,')
\right]{\phi}_{\vec{n},k,\sigma}(\vec{r}\,)  \nonumber\\
&&-\int d\vec{r}\,'V(\vec{r}-\vec{r}\,')
{\phi}_{\vec{n},k,\sigma}(\vec{r}\,')
\sum_{\vec{m},p}\nu_{\vec{m},p,\sigma}
{\phi}_{\vec{m},p,\sigma}{}^\dagger(\vec{r}\,')
{\phi}_{\vec{m},p,\sigma}(\vec{r}\,),
\label{HF_equation_general}
\end{eqnarray}
where $\nu_{\vec{m},p,s}$ 
is the filling factor at the specific quantum number, and
it satisfies
\begin{eqnarray}
N_e=\sum_{\vec{m},p,\sigma}\nu_{\vec{m},p,\sigma},
\end{eqnarray}
where $N_e$ is the total electron number. ${\cal H}_0$ is the same as 
in Eq. (\ref{noninteracting_H}) by taking $p_y=k$.
Note that the positive charge donor density (which produces 
the electron gas and thus provides charge neutrality for the 
whole system) is not explicitly included above
because these donors are usually located far away
from the well in the experiment. In general, this 
background doping effect can be  
effectively included by introducing a screening length, $\lambda$,
into the bare Coulomb interaction, $V(\vec{q}\,)$, by writing
$V(\vec{q}\,)=(4\pi e^2/\epsilon_0)(|\vec{q}\,|^2+(2\pi/\lambda)^{2})^{-1/2}$. 
We take $\lambda=620$ \AA$\ $ in our numerical calculation below 
to be comparable to the experimental setting \cite{pan01}.
This regularization of Coulomb interaction has little quantitative 
or qualitative effects on the results shown in this paper.
The details of the donor screening and the exact value of $\lambda$
do not in any way affect any of our qualitative conclusions.
For the situation we focus in this paper,
electrons are assumed to be uniformly distributed in the 2D well plane
(i.e. $\nu_{\vec{m},p,s}$ is independent of guiding center coordinate,
$p$), and therefore Eq. (\ref{HF_equation_general}) can be simplified
further as shown in Appendix \ref{HF_equation_append}. 
To solve the SCHF equation, we first use
the noninteracting wavefunction to calculate the HF matrix elements and then
diagonalize it to get new eigenstates, which are used to calculate the
HF matrix element again iteratively until self-consistency is achieved.
The new single electron Green's function
in SCHFA is similar to the noninteracting
one in Eq. (\ref{Greenfunc2}) except that the wavefunctions 
and energies correspond to the Hartree-Fock theory:
\begin{eqnarray}
G_{\vec{n},\sigma}(\vec{r}_1,\vec{r}_2;\omega)&=&
\frac{\sum_{k}{\phi}_{\vec{n},k,\sigma}{}^\dagger(\vec{r}_2)
{\phi}_{\vec{n},k,\sigma}(\vec{r}_1)}
{\left[{\cal G}^0_{\vec{n},\sigma}(\omega)\right]^{-1}-
\Sigma_{\vec{n},\sigma}^{HF}}
=\sum_{k}{\phi}_{\vec{n},k,\sigma}{}^\dagger(\vec{r}_2)
{\phi}_{\vec{n},k,\sigma}
(\vec{r}_1)\cdot {\cal G}_{\vec{n},\sigma}(\omega)
\label{G_w}
\end{eqnarray}
where ${\cal G}_{\vec{n},\sigma}^{0}(\omega)$ 
is Fourier transform of the noninteracting propagator,
${\cal G}_{\vec{n},\sigma}^{0}(t)$:
\begin{eqnarray}
{\cal G}_{\vec{n},\sigma}^{0}(\omega)=
\int dt\,e^{i\omega t}{\cal G}_{\vec{n},\sigma}^0 (t)=\frac{1}{i\omega-
E^0_{\vec{n},\sigma}-\mu},
\end{eqnarray}
and similarly
\begin{eqnarray}
{\cal G}_{\vec{n},\sigma}(\omega)=
\frac{1}{i\omega-
E^0_{\vec{n},\sigma}-\Sigma_{\vec{n},\sigma}^{HF}-\mu},
\end{eqnarray} 
where the Hartree-Fock self-energy, $\Sigma_{\vec{n},\sigma}^{HF}=
{E}_{\vec{n},\sigma}-E^0_{\vec{n},\sigma}$, is obtained
from the self-consistent solution of the Hartree-Fock problem 
(see Appendix \ref{HF_equation_append}, in particular 
Eq. (\ref{selfconsistent_HF})). 

\subsection{Level Crossing and Total Energy}
\label{level_crossing}

\begin{figure}
 \vbox to 8cm {\vss\hbox to 5.5cm
 {\hss\
    {\includegraphics{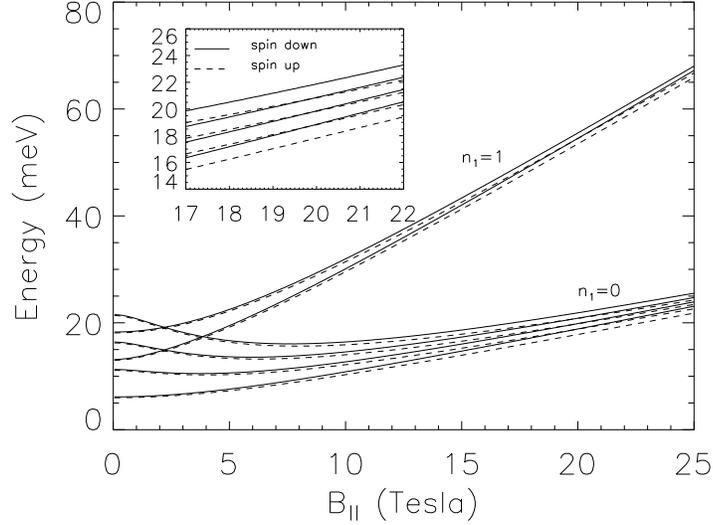}
  }
  \hss}
  }
\caption{
Calculated landau level energy spectra for 
noninteracting electrons in a parabolic quantum well
with a parallel (in-plane) magnetic field, $B_\|$. 
The system parameters are chosen to be
the same as the experimental data in ref. [7] for $\nu=6$.
}
\label{energy_levels_figure}
\end{figure}

For our purposes, the most important feature of the spectra shown in Fig.
\ref{energy_levels_figure} is the existence of a single particle level crossing
at $B_{\|}=19.8$ Tesla (for the noninteracting system), where
$\omega_1\sim 36$ meV, and $\omega_2=\omega_z\sim 1$ meV.
The origin of this
crossing can be understood by considering the asymptotic form
of the energy levels in Eq. (\ref{omega12}) for large parallel
magnetic fields: $\omega_2 \rightarrow \omega_0\omega_\bot/\omega_{\|}
\rightarrow 0$, so that at a critical in-plane magnetic field
value ($B_\|^\ast$) $\omega_2$
becomes smaller than the Zeeman energy leading to the level crossing 
shown in Fig. \ref{energy_levels_figure}.
For a non-interacting
electron gas this level crossing leads necessarily to a (rather trivial)
first order phase transition at $B_\|=B_\|^\ast$
with an abrupt change in spin polarization for systems
at even filling factors.
Interesting quantum phase transition that may take 
place around this level crossing
is the main subject of this article.
In particular, we wish to investigate whether quantum level repulsion 
converts this first order transition to a second order quantum phase 
transition around this degeneracy point.
Our calculated mode dispersion can be directly compared to inelastic
light scattering spectroscopy, when such experiments are eventually 
carried out in these WWW systems in tilted fields. One can easily
use our analytical results to study the magnetoplasmon mode dispersion
at odd integer filling factors or for weaker in-plane magnetic
field values (where intersubband coupling needs to be included).

By using the self-energy obtained from Eq. (\ref{HF_equation_general}) in SCHFA,
the total energy of an interacting
quantum Hall system can be obtained for a given electron configuration
$(\vec{N}_\uparrow,\vec{N}_\downarrow)$, where $\vec{N}_\sigma$ is the orbital
level index of the highest filled level of spin $\sigma$. Considering 
double counting of interaction energy, the total energy in HF
approximation is
\begin{eqnarray}
E^{HF}_{tot}(\vec{N}_\uparrow,\vec{N}_\downarrow)&=&
\sum_{\sigma}\sum_{E^0_{\vec{n},\sigma}\leq E^0_{\vec{N}_\sigma,\sigma}}
\left[E^0_{\vec{n},\sigma}+
\frac{1}{2}\Sigma^{HF}_{\vec{n},\sigma}\right].
\label{total_energy_HF}
\end{eqnarray}
To obtain the ground state energy, $E_G$, one should compare the total
energies of all possible electron configurations and determine which
one gives the lowest energy. As indicated in Fig.
\ref{energy_levels_figure},
a first order (noninteracting) phase transition from an unpolarized
ground state (i.e. $\vec{N}_\uparrow=\vec{N}_\downarrow=(0,\nu/2)$) 
to a polarized
ground state ($\vec{N}_\downarrow=\vec{N}_\uparrow+(0,2)$) is expected to
happen at a critical in-plane magnetic field, $B_\|^\ast$.
In the third column of Table \ref{critical_B_table} we show our
numerical calculation results of $B_\|^\ast$ obtained from Eq.
(\ref{total_energy_HF}) in the first order HF approximation
for even filling factors $\nu=6,8$.
When the total electron density is fixed, $B_\|^\ast$ is larger as the filling
factor, $\nu$, is lowered by increasing the perpendicular magnetic field.
Therefore our HF results qualitatively agree with the experimental data 
presented in Ref. \cite{pan01} except for a lower estimate of the
critical magnetic field, $B_\|^\ast$, which may be due to the correlation
effects not included in the HF approximation and/or the nonparabolicity
of the realistic confinement potential of the quantum well sample used
in Ref. \cite{pan01}. 

\section{Magnetoplasmon Excitations}
\label{magnetoplasmon}

In this section we will develop the full theory of magnetoplasmon
excitations of an integer quantum Hall system 
confined in a parabolic well
and subject to a tilted magnetic field within TDHFA.
For zero width (pure 2D) wells with a perpendicular magnetic field only,
magnetoplasmon modes were investigated in \cite{kallin}.
We note that magnetoplasmon excitations
in parabolic wells have been theoretically discussed previously
in the literature
\cite{brey89,brey89_sdw,brey91_cdw,dempsey,marmorkos93,stanescu00} 
in different limited conditions. Our work goes
beyond results presented in those papers and we derive the exact 
dispersion of collective modes in the lowest order
of the ratio of Coulomb interaction to the noninteracting 
Landau level separation. 
In a WWW with tilted magnetic field, 
there is no translational symmetry along the growth
direction ($z$), which is hybridized with 
the in-plane components ($x-y$)
so that a many-body theory developed in momentum space seems not 
to be particularly useful. However, it is shown
below in Sec. \ref{dipole} that
the in-plane momentum of an electron-hole dipole in such WWW
with tilted magnetic field is still conserved, showing the existence of  
a well-defined electron-hole bound state (a
{\it magnetic exciton}) \cite{kallin} and the collective mode
dispersion along the 2D plane can still be obtained analytically
as we show below.
The full many-body theory and the numerical results for collective mode
dispersion are shown in Secs. \ref{correlation_function}-
\ref{pl_disperion} and in Sec. \ref{plasmon_numerical} respectively.

\subsection{Momentum Conservation of an Electron-hole Dipole Pair}
\label{dipole}

As pointed out in Ref. \cite{kallin}, a crucial fact that
allows one to explicitly write analytical
expressions of the energy dispersion of magnetoplasmon excitations is 
the existence of a good quantum number in the problem given by the
well defined in-plane momentum of the electron-hole dipole pair
(magnetic exciton).
It is easy to show that their argument can be extended to 
the case of a WWW with an {\it arbitrary} confinement potential
along the $z$ direction even in the presence of a tilted magnetic field. This
is not obvious since the tilted
magnetic field typically hybridizes the in-plane motion
with the subband dynamics perpendicular to the plane, destroying the apparent
translational symmetry.
Consider the Hamiltonian of 
a magnetic exciton or an electron-hole pair in a general quasi-2D system,
\begin{eqnarray}
{\cal H}_{X} = \frac{1}{2m^*}
\left[ (\vec{ p}_1 - \frac{e}{c}\vec{A}(\vec{r}_1))^2
+ (\vec{p}_2 + \frac{e}{c}\vec{A}(\vec{r}_2))^2 \right]
-V(\vec{r}_1- \vec{r}_2) + U(z_1)+U(z_2),
\end{eqnarray}
where particle momenta $\vec{p}_i$, vector potential $\vec{A}$,
and particle coordinates  $\vec{r}_i$ 
are all three dimensional vectors. $V(\vec{r})$ and $U(z)$ are 
electron-electron Coulomb interaction and the quantum well confinement
potential respectively. The
Zeeman term is neglected here because it is irrelevant for this discussion.
Following \cite{kallin} a magnetic exciton momentum operator can be defined
to be
\begin{eqnarray}
\vec{Q}_{X} = \vec{p}_1 +  \vec{p}_2
-\frac{e}{c} (  \vec{A}(\vec{r}_1) -  \vec{A}(\vec{r}_2) )
+\frac{e}{c}  \vec{B}_{tot}\times (  \vec{r}_1 -  \vec{r}_2 ),
\label{diple_momentum}
\end{eqnarray}
where $\vec{B}_{tot}=B_\perp\hat{z}+B_\|\hat{x}$.
Using the Landau gauge for the vector potential 
one can easily verify that the in-plane components $\vec{Q}_{X,\perp}=
(Q_{X,x},Q_{X,y})$ 
commute with the Hamiltonian. Existence of dipole excitations with
well defined momenta (eigenvalues of $\vec{Q}_{X,\perp}$)
immediately follows from this commutation.  
Similar to Ref. \cite{kallin}, we can construct 
the zero-momentum magnetic exciton wavefunction in a parabolic well
with a tilted magnetic field:
\begin{eqnarray}
\Psi^{\sigma_\beta,\sigma_\alpha}
_{\vec{n}_\beta,\,\vec{n}_\alpha}(\Delta x,\Delta y,Z,\Delta z)
=\int d\eta\, e^{-i\eta\Delta y/l_0^2}
\Phi_{\vec{n}_\beta,\sigma_\beta}(\eta+\Delta x/2,Z+\Delta z/2)
\Phi_{\vec{n}_\alpha,\sigma_\alpha}(\eta-\Delta x/2,Z-\Delta z/2),
\label{wavefunction-WW}
\end{eqnarray}
where a hole is in a state $\vec{n}_\alpha=\vec{n}$ and
a particle is in a state $\vec{n}_\beta
=\vec{n}_\alpha+\vec{m}$. For the exciton wavefunction of 
finite momentum, $\vec{q}_\perp$, one just needs to replace
$\Delta\vec{r}_\perp=(\Delta x,\Delta y)$ by $\Delta\vec{r}_\perp-l_0^2
\vec{q}_\perp\times\hat{z}$, and introduce a plane wave prefactor for the 
center of mass coordinate \cite{kallin}.
Note that this wavefunction has additional dynamics along $z$ direction:
center of mass ($Z$) and relative
($\Delta z$) coordinates of the electron-hole pair.
Other than this additional $z$ dynamics,
the only difference between the exciton wavefunctions 
for the ZWW 2D system \cite{kallin} with a perpendicular magnetic
field (see Eq. (\ref{wavefunctions-KH}) or \cite{kallin}) 
and for the WWW with a tilted
magnetic field (our interest in this paper) is that the latter 
has one more Landau level
quantum number associated with the subband dynamics induced by the 
confinement energy of the well. 
In Appendix \ref{realspace_expression} we show that
the magnetoplasmon energy in our theory can be expressed
in terms of the magnetic exciton wavefunction given in Eq. 
(\ref{wavefunction-WW}). This provides a more comprehensive 
and physical picture for understanding
the collective mode excitations discussed in this paper
(see Appendix \ref{realspace_expression} in this context).

\subsection{Correlation Function}
\label{correlation_function}

In the linear response theory, collective mode energies
are obtained by the poles of a density correlation 
function, $\Pi_\lambda$, where $\lambda=\rho,\ S_\pm$ and $S_z$,
for the singlet charge density mode and the three triplet spin density modes
respectively.
We define an operator, $\Theta_\lambda=1,\ 2S_\pm,\ 2S_z$, respectively
for the spin vertex operator of each corresponding correlation function.
In this notation the most general form of these correlation 
functions in coordinate
space is \cite{fetter}
\begin{eqnarray}
\Pi_\lambda(\vec{r},t;\vec{r}\,',t')&=&-i\sum_{\sigma_{1,2}}
\sum_{\sigma_{1,2}'}[\Theta_\lambda ]_{\sigma_1,\sigma_2}
[\Theta_\lambda ]_{\sigma_1',\sigma_2'}
\left\langle T
\left[\hat{\Psi}^\dagger_{\sigma_1}(\vec{r},t)
\hat{\Psi}^{}_{\sigma_2}(\vec{r},t)
\hat{\Psi}^\dagger_{\sigma_1'}(\vec{r}\,',t')
\hat{\Psi}^{}_{\sigma_2'}(\vec{r}\,',t')\right]\right\rangle_G,
\label{response_func}
\end{eqnarray}
where $\hat{\Psi}^\dagger_{\sigma}(\vec{r},t)
(\hat{\Psi}_{\sigma}(\vec{r},t))$ are the electron field creation (annihilation)
operators of space $\vec{r}$ and spin $\sigma$ at time $t$;
$T[\cdots]$ is time-order operator, and 
$\langle\cdots\rangle_G$ is the expectation value of the
interacting ground state. In a WWW system,  
there is no translational symmetry along $z$ direction so that one 
has no correlation function in momentum space in $z$ direction.
The usual momentum space description for the vertex 
function and the related Dyson's equation then seems
not feasible
because the in-plane magnetic field mixes the $z$ dynamics 
with in-plane dynamics \cite{brey89_sdw,brey91_cdw,marmorkos93}. 
Actually the system is more like a 2D quantum dot \cite{yip}
in $x-z$ plane confined 
by two independent parabolic potentials along ${\bar{x}}$ and
${\bar{z}}$ axes as shown in Eq. (\ref{Hamiltonian_rotated}).
The method we develop in this paper, however, enables one to obtain directly
the appropriate Dyson's equations for the screened interaction
and the vertex function without evaluating the correlation
functions of Eq. (\ref{response_func}).
The magnetoplasmon excitation dispersion and the dielectric function relevant
for screening
can be read out directly from our equations given in the next section.
Note that the theory developed below is independent of the exact form of
the single electron wavefunction and is completely general within
the TDHFA.

\subsection{Screened Interaction and Vertex Function} 

Before exploring the many-body theory for the collective mode, 
we first define the interaction matrix element, which will be 
used frequently later.
Using the interacting single particle wavefunction, the
unscreened matrix element of a bare Coulomb interaction, $V(\vec{r}\,)$,
can be obtained
\begin{eqnarray}
V^{k_1k_4,k_2,k_3,\sigma_1\sigma_2}_{\vec{n}_1\vec{n}_4,\vec{n}_2\vec{n}_3}
\delta_{\sigma_1\sigma_4}\delta_{\sigma_2\sigma_3}
&=&\int d\vec{r}_1\int d\vec{r}_2
\,V(\vec{r}_1-\vec{r}_2){\phi_{\vec{n}_1,k_1,\sigma_1}}^\dagger(\vec{r}_1)
{\phi_{\vec{n}_2,k_2,\sigma_2}}^\dagger(\vec{r}_2)
\phi_{\vec{n}_3,k_3,\sigma_3}(\vec{r}_2)
\phi_{\vec{n}_4,k_4,\sigma_4}(\vec{r}_1)
\delta_{\sigma_1\sigma_4}\delta_{\sigma_2\sigma_3}
\nonumber\\
&=&\frac{1}{\Omega}\sum_{\vec{q}}\delta_{k_4-k_1,-q_y}\delta_{k_3-k_2,q_y}
e^{-i(k_1-k_2-q_y)q_xl_0^2}
V^{\sigma_1\sigma_2}_{\vec{n}_1\vec{n}_4,\vec{n}_2\vec{n}_3}(\vec{q}\,)
\delta_{\sigma_1\sigma_4}\delta_{\sigma_2\sigma_3},
\label{bare_matrix}
\end{eqnarray}
where $\Omega$ is the well volume and we
define an effective interaction, $V^{\sigma_1\sigma_2}
_{\vec{n}_1\vec{n}_4,\vec{n}_2\vec{n}_3}
(\vec{q}\,)\equiv V(\vec{q}\,)
A^{\sigma_1\sigma_1}_{\vec{n}_1\vec{n}_4}(-\vec{q}\,)
A^{\sigma_2\sigma_2}_{\vec{n}_2\vec{n}_3}(\vec{q}\,)$, where
the form factor $A^{\sigma_i\sigma_j}_{\vec{n}_i\vec{n}_j}(\vec{q}\,)$ is
obtained from the single particle wavefunctions
\begin{eqnarray}
A^{\sigma_i\sigma_j}
_{\vec{n}_i\vec{n}_j}(\vec{q}\,)&=&\int d\vec{r}\,e^{-i\vec{q}\cdot\vec{r}}
{\phi_{\vec{n}_i,-q_y/2,\sigma_i}}^\dagger(\vec{r}\,)
{\phi_{\vec{n}_j,q_y/2,\sigma_j}}(\vec{r}\,)
\nonumber\\
&=&\int dx\int dz\,e^{-iq_xx-iq_zz}
\Phi_{\vec{n}_i,\sigma_i}(x-q_yl_0^2/2,z)
\Phi_{\vec{n}_j,\sigma_j}(x+q_yl_0^2/2,z).
\label{A_def}
\end{eqnarray}
Momentum and spin conservations during the scattering process
have been included in Eq. (\ref{bare_matrix}).

\begin{figure}
 \vbox to 8cm {\vss\hbox to 5.5cm
 {\hss\
   {\includegraphics{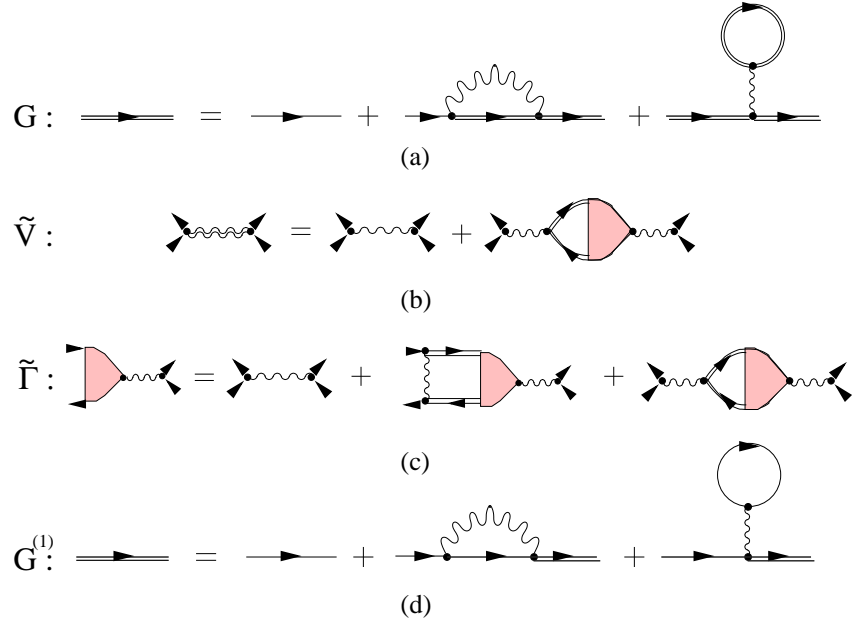}
   }
  \hss}
 }
\caption{
Feynman diagrams of the time-dependent Hartree-Fock
approximation. Solid lines are single particle
Green's function and wavy lines are Coulomb interaction.
Single(double) lines are bare(dressed) Green's function and/or interaction:
(a) the self-consistent Hartree-Fock approximation for the 
single electron Green's function;
(b) and (c) are respectively the Dyson's equations for
electron-electron interaction, single electron Green's function
and vertex function
in the time-dependent Hartree-Fock approximation.
The second term of (c) is the ladder series, while the third term is
the bubble series (RPA diagram), which does not appear 
when calculating the vertex function
for spin-flip excitations (since the interaction is spin-conserving)
as mentioned in the text.
(d) is the Green's function in the first order Hartree-Fock approximation.
}
\label{feynm_figure}
\end{figure}

Note that in the strong parallel magnetic field regime,
(10 Tesla $<B_\|<25$ Tesla), only a few Landau levels of the first subband
($n_1=0$) are occupied at zero temperature since $\omega_1\gg\omega_2$
(see Eq. (\ref{energy_levels_eqn}) and Fig. \ref{energy_levels_figure}). 
Therefore we could omit the
first orbital (i.e. subband) level index and neglect
all intersubband transitions
(i.e. excitations between levels of different $n_1$)
by assuming for simplicity that $n_1=0$ throughout our analysis and
numerical calculations shown below except where
noted otherwise. In other words, the vector representation used
in Eq. (\ref{omega12}) for orbital Landau level index, 
$\vec{n}=(n_1,n_2)=(0,n_2)$, is simplified to be $n$ and so are all
other orbital notations (like $\vec{m}=(0,m_2)=m$ and $\vec{N}_\sigma
=(0,N_\sigma)=N_\sigma$, etc.) from now on in this paper.
It is straightforward to extend all of our analytical and numerical results
to include excitations of both orbital quantum numbers.
All analytical results would retain the same form with
additional level indices (i.e. other value of $n_1$) showing up
in the formula. Our
numerical results will not be affected at all by this assumption
in the strong in-plane
magnetic field region of our interest where hybridization with higher
$n_1$ levels is negligiblly small.
We also will not show the spin index explicitly during the derivation 
except in the final results.
We start from the screened Coulomb interaction, $\tilde{V}(\vec{r}_1,
\vec{r}_2,t_1-t_2)$, caused by electron-hole polarization
(see Fig. \ref{feynm_figure}(b)):
\begin{eqnarray}
\tilde{V}(\vec{r}_1,\vec{r}_2;t_1-t_2)
&=&V(\vec{r}_1-\vec{r}_2)
\delta(t_1-t_2)+\int d\vec{r}_3\int d\vec{r}_4 V(\vec{r}_1-\vec{r}_3)
\Pi(\vec{r}_3,t_1;\vec{r}_4,t_2)V(\vec{r}_4-\vec{r}_2),
\label{RPA_realspace}
\end{eqnarray}
where $\Pi(\vec{r}_1,t_1;\vec{r}_2,t_2)=\Pi_\rho(\vec{r}_1,t_1;\vec{r}_2,t_2)$
is the reducible charge polarizability (see Eq. (\ref{response_func})).
Multiplying by single particle wavefunctions and doing the space integration,
Eq. (\ref{RPA_realspace}) can be transformed to
\begin{eqnarray}
\tilde{V}_{1,4;2,3}(t_1-t_2)&\equiv&\int d\vec{r}_1\int d\vec{r}_2
\,\tilde{V}(\vec{r}_1,\vec{r}_2;t_1-t_2)
{\phi_{1}}^\dagger(\vec{r}_1){\phi_{2}}^\dagger(\vec{r}_2)
\phi_{3}(\vec{r}_2)\phi_{4}(\vec{r}_1)
\nonumber\\
&=&V_{1,4;2,3}\delta(t_1-t_2)
+\sum_{\alpha\beta}V_{1,4;\beta\alpha}
\left[\int dt_5 dt_6 {\cal G}_{\alpha}(t_1-t_5){\cal G}_{\beta}(t_6-t_1)\right.
\nonumber\\
&&\hspace{1cm}\times\left.
\int_{2,4,5,6}{\phi_{\alpha}}^\dagger(\vec{r}_5)\phi_{\beta}(\vec{r}_6)
\gamma(\vec{r}_5,t_5;\vec{r}_6,t_6;\vec{r}_4,t_2)
V(\vec{r}_4-\vec{r}_2)
{\phi_{2}}^\dagger(\vec{r}_2)\phi_{3}(\vec{r}_2)
\right] 
\nonumber\\
&=&
V_{1,4;2,3}\delta(t_1-t_2)
+\sum_{\alpha\beta}V_{1,4;\beta\alpha}
\tilde{\Gamma}_{\alpha\beta;2,3}(t_1-t_2),
\label{chain_newspace}
\end{eqnarray}
where we have introduced a conventional reducible vertex function, 
$\gamma(\vec{r}_5,t_5;\vec{r}_6,t_6;\vec{r}_4,t_2)$, in coordinate space 
to express 
the reducible polarizability, $\Pi$; index
$\alpha(\beta)$ denotes all related quantum numbers of that level, 
$(m_{\alpha(\beta)},p_{\alpha(\beta)},\sigma_{\alpha(\beta)})$, i.e.
$\phi_\alpha(\vec{r})=\phi_{m_{\alpha},p_{\alpha},\sigma_{\alpha}}(\vec{r})$,
${\cal G}_\alpha(t)={\cal G}_{m_\alpha,\sigma_{\alpha}}(t)$, 
$V_{1,4;\alpha\beta}
=V^{k_1k_4,p_\alpha p_\beta,\sigma_1\sigma_\alpha}
_{n_1n_4,m_\alpha m_\beta}$, and
$\tilde{\Gamma}_{\alpha\beta;2,3}=\tilde{\Gamma}^{p_\alpha p_\beta;
k_2k_3;\sigma_\alpha\sigma_\beta,\sigma_2\sigma_3}_{m_\alpha m_\beta;n_2n_3}$ 
for simplicity (number indices represent
external variables, 
while Greek indices represent dummy variables in a summation) and
$\int_i\equiv\int d\vec{r}_i$.
To avoid confusion, we clarify our notations which are necessarily different
from the standard many-body textbook terminology because of the highly
complicated nature of our single-particle wavefunctions.
First $V_{1,4;2,3}$ and $V_{n_1n_4,n_2n_3}(\vec{q})$
are different functions according to their definition 
in Eq. (\ref{bare_matrix}); secondly the $\tilde{\Gamma}$ function
in Eq. (\ref{chain_newspace}) is not the same as the 
conventional definition of a vertex 
function due to our inclusion in $\tilde{\Gamma}$ of additional 
two electron Green's functions
and one interaction term (however, $\gamma$ is the same as the conventional 
reducible vertex function in coordinate space. See Fig. \ref{feynm_figure}
(c)). This is because, unlike a ZWW
(pure 2D) in Ref. \cite{kallin} or a WWW without any
in-plane magnetic field \cite{brey91_cdw}, the $z$-component of electron
wavefunction of our system is not separable and therefore cannot be
ignored. It is more
convenient to work in the relevant conserved quantum number 
space rather than
in the conventional momentum space. 

The leading order (of the ratio of the interaction strength to the 
noninteracting energy separation, $\omega_2$) of the vertex 
function, $\tilde{\Gamma}_{\alpha\beta;2,3}(t_1-t_2)$,
is obtained by using $\gamma(\vec{r}_5,t_5;\vec{r}_6,t_6;\vec{r}_4,t_2)=
\delta(\vec{r}_5-\vec{r}_4)\delta(\vec{r}_6-\vec{r}_4)\delta(t_5-t_2)
\delta(t_6-t_2)$ in Eq. (\ref{chain_newspace}):
\begin{eqnarray}
\Gamma_{\alpha\beta;2,3}(t_1-t_2)
&=&{\cal G}_{\alpha}(t_1-t_2){\cal G}_{\beta}(t_2-t_1)V
_{\alpha \beta;2,3},
\end{eqnarray}
which has the following Fourier transform in time:
\begin{eqnarray}
\Gamma_{\alpha\beta;2,3}(\omega)=D_{\alpha\beta}(\omega)
V_{\alpha \beta;2,3},
\label{bare_vertex}
\end{eqnarray}
where (after retrieving the spin index)
\begin{eqnarray}
D_{\alpha\beta}(\omega)&=&
\frac{\theta(m_{\sigma_\alpha}-N_{\sigma_\alpha})
\theta(N_{\sigma_\beta}-m_{\sigma_\beta})
-\theta(m_{\sigma_\beta}-N_{\sigma_\beta})
\theta(N_{\sigma_\alpha}-m_{\sigma_\alpha})}
{(m_{\sigma_\beta}-m_{\sigma_\alpha})\omega_2-
(\sigma_\beta-\sigma_\alpha)\omega_z+i\omega},
\label{D_w}
\end{eqnarray}
is nonzero only when the dipole pair, $(\alpha,\beta)$, represents
one hole in the filled level and one
electron in the empty level at zero temperature.
To avoid confusion, here we clarify the meaning of $\sigma_{\alpha(\beta)}$,
$m_{\sigma_{\alpha(\beta)}}$,
and $N_{\sigma_{\alpha(\beta)}}$ in Eq. (\ref{D_w}) again: 
$\sigma_{\alpha(\beta)}$ is the spin quantum number of state $\alpha(\beta)$,
$m_{\sigma_{\alpha(\beta)}}=(0,m_{\sigma_{\alpha(\beta)}})=
\vec{m}_{\sigma_{\alpha(\beta)}}$ is the orbital Landau level index
of state $\alpha(\beta)$, and $N_{\sigma_{\alpha(\beta)}}=
(0,N_{\sigma_{\alpha(\beta)}})=\vec{N}_{\sigma_{\alpha(\beta)}}$ is the
orbital Landau level index of the highest filled level of spin 
$\sigma_{\alpha(\beta)}$ as first defined in Eq. (\ref{total_energy_HF}).
We will see later that $D_{\alpha\beta}(\omega)$ is the only
dynamical part of the vacuum electron-hole bubble. 
Note that when spin is included, $V_{\alpha\beta,\gamma\lambda}$
implies $\sigma_{\alpha}=\sigma_\beta$ and 
$\sigma_\gamma=\sigma_\lambda$ automatically because of
the manifestly spin-conserving non-spin-flip nature of
Coulomb interaction. As a consequence, $\Gamma_{\alpha\beta;2,3}(t_1-t_2)$ in
Eq. (\ref{bare_vertex})
(but not $\gamma(\vec{r}_5,t_5;\vec{r}_6,t_6;\vec{r}_4,t_2)$ in 
Eq. (\ref{chain_newspace})) then becomes identically zero when 
considering spin-flip
excitations.
\subsection{Dyson's Equations in TDHFA} 

Including ladder and bubble diagrams as shown
in Fig. \ref{feynm_figure}(c), Dyson's equation for the full vertex function,
$\tilde{\Gamma}_{\alpha\beta;2,3}(t_1-t_2)$, is
\begin{eqnarray}
&&\tilde{\Gamma}_{\alpha\beta;2,3}(t_1-t_2)
={\Gamma}_{\alpha\beta;2,3}(t_1-t_2)
+\int dt_5 dt_6 {\cal G}_{\alpha}(t_1-t_5)
{\cal G}_{\beta}(t_6-t_1)\delta(t_5-t_6)
\nonumber\\
&&\times
\int_{2,4,5,6}{\phi_{\alpha}}^\dagger(\vec{r}_5)\phi_{\beta}(\vec{r}_6)
V(\vec{r}_4-\vec{r}_2)
{\phi_{2}}^\dagger(\vec{r}_2)\phi{3}(\vec{r}_2)
\left[-\int_{7,8}\int dt_7 dt_8
G(5,7)G(8,6)
V(\vec{r}_5-\vec{r}_6)
\gamma(\vec{r}_7,t_7;\vec{r}_8,t_8;\vec{r}_4,t_2)\right]
\nonumber\\
&&+\int dt_5 dt_6 {\cal G}_{\alpha}(t_1-t_5)
{\cal G}_{\beta}(t_5-t_1)\delta(t_5-t_6)
\nonumber\\
&&\times
\int_{2,4,5,6}{\phi_{\alpha}}^\dagger(\vec{r}_5)\phi_{\beta}(\vec{r}_5)
V(\vec{r}_4-\vec{r}_2)
{\phi_{2}}^\dagger(\vec{r}_2)\phi_3(\vec{r}_2)
\left[\int_{7,8}\int dt_7 dt_8
G(6,7)G(8,6)V(\vec{r}_5-\vec{r}_6)
\gamma(\vec{r}_7,t_7;\vec{r}_8,t_8;\vec{r}_4,t_2)\right],
\label{dynson1}
\end{eqnarray}
which can be further simplified to 
\begin{eqnarray}
&&\tilde{\Gamma}_{\alpha\beta;2,3}(t_1-t_2)
={\Gamma}_{\alpha\beta;2,3}(t_1-t_2)
+\int dt_5 {\cal G}_{\alpha}(t_1-t_5){\cal G}_{\beta}(t_5-t_1)\sum_{\mu\nu}
\left[-V_{\alpha\mu;\nu\beta}+V_{\alpha\beta;\nu\mu}\right]
\tilde{\Gamma}_{\mu\nu;2,3}(t_5-t_2),
\end{eqnarray}
with the following Fourier transform in time 
\begin{eqnarray}
\tilde{\Gamma}_{\alpha\beta;2,3}(\omega)
={\Gamma}_{\alpha\beta;2,3}(\omega)
+D_{\alpha\beta}(\omega)\sum_{\mu\nu}
\left[-V_{\alpha\mu;\nu\beta}+V_{\alpha\beta;\nu\mu}\right]
\tilde{\Gamma}_{\mu\nu;2,3}(\omega).
\label{dynson_Gamma}
\end{eqnarray}
Similarly Fourier transform of Eq. (\ref{chain_newspace}) gives
\begin{eqnarray}
\tilde{V}_{1,4;2,3}(\omega)&=&V_{1,4;2,3}+\sum_{\alpha\beta}V_{1,4;\beta\alpha}
\tilde{\Gamma}_{\alpha\beta;2,3}(\omega).
\label{dynson_V}
\end{eqnarray}
Eq. (\ref{dynson_Gamma}) and Eq. (\ref{dynson_V}) are respectively
Dyson's equations
for the vertex function and the interaction matrix element in the quantum
number, $\alpha\equiv(m_\alpha,p_\alpha,\sigma_\alpha)$, space.

In order to investigate the magnetoplasmon dispersion, one has 
to integrate out the continuous variable, $k$, in 
Eq. (\ref{dynson_Gamma}) and Eq. (\ref{dynson_V}) to get a matrix 
representation in the
level index only. Taking into account the momentum conservation shown
in Eq. (\ref{bare_matrix}), we define a new unscreened matrix
element and a new bare vertex function given by (let $\vec{q}_\perp\equiv
q_x\hat{x}+q_y\hat{y}$ be the in-plane momentum)
\begin{eqnarray}
U_{n_1n_4,n_2n_3}(\vec{q}_\perp)&\equiv&
\sum_{k_1} e^{i(k_1-k_2)q_xl_0^2}\,V_{n_1n_4,n_2n_3}
^{k_1+q_y/2,k_1-q_y/2;k_2-q_y/2,k_2+q_y/2}
=\frac{1}{2\pi l_0^2 L_z}\sum_{q_z'}V_{n_1n_4,n_2n_3}(\vec{q}_\perp,q_z')
\label{U_def}
\\
\Lambda_{m_\alpha m_\beta;n_2n_3}(\vec{q}_\perp,\omega)&\equiv&
[D_{m_\alpha m_\beta}(\omega)]^{-1}
\sum_{k_1} e^{i(k_1-k_2)q_xl_0^2}\,{\Gamma}^
{k_1+q_y/2,k_1-q_y/2;k_2-q_y/2,k_2+q_y/2}_{m_\alpha m_\beta;n_2n_3}
=U_{m_\alpha m_\beta;n_2n_3}(\vec{q}_\perp),
\label{Lambda}
\end{eqnarray}
where $L_xL_y=2\pi l_0^2 N_\phi$, and $N_\phi$ is the degeneracy
of each Landau level. Note that in Eqs. (\ref{U_def}) and (\ref{Lambda})
only the in-plane component of momentum, $\vec{q}_\perp$, is shown explicitly.
This follows from the fact that the in-plane exciton momentum is a good quantum
number even in the presence of tilted magnetic field as discussed in Sec.
\ref{dipole}.
Expressions for the screened matrix element, $\tilde{U}$, 
and the full vertex function, $\tilde{\Lambda}$, 
can be similarly obtained using the $k$-summation over
$\tilde{V}$ and $\tilde{\Gamma}$ as in Eqs. (\ref{U_def}) and (\ref{Lambda}).
After some tedious analysis, we obtain the following pair of 
matrix equations in Landau level indices 
\begin{eqnarray}
\tilde{U}_{n_1n_4,n_2n_3}(\vec{q}_\perp,\omega)&=&
U_{n_1n_4,n_2n_3}(\vec{q}_\perp)
+\sum_{m_\alpha m_\beta}U_{n_1n_4;m_\beta m_\alpha}(\vec{q}_\perp)
D_{m_\alpha m_\beta}(\omega)
\,\tilde{\Lambda}_{m_\alpha m_\beta;n_2n_3}(\vec{q}_\perp,\omega),
\label{dynson_U2}
\\
\tilde{\Lambda}_{m_\alpha m_\beta;n_2n_3}(\vec{q}_\perp,\omega)
&=&\Lambda_{m_\alpha m_\beta;n_2n_3}(\vec{q}_\perp,\omega)
+\sum_{m_\mu m_\nu}
W_{m_\alpha m_\beta;m_\nu m_\mu}(\vec{q}_\perp)
D_{m_\mu m_\nu}(\omega)
\tilde{\Lambda}_{m_\mu m_\nu;n_2n_3}(\vec{q}_\perp,\omega),
\label{dynson_Lambda2}
\end{eqnarray}
where the new interaction function, $W$, is 
\begin{eqnarray}
W_{m_\alpha m_\beta;m_\nu m_\mu}(\vec{q}_\perp)\equiv
-U^{bind}_{m_\alpha m_\mu;m_\nu m_\beta}(\vec{q}_\perp)
+U_{m_\alpha m_\beta;m_\nu m_\mu}(\vec{q}_\perp).
\label{W_def}
\end{eqnarray}
The ladder (exciton binding) energy, $U^{bind}(\vec{q}_\perp)$,
and the random-phase-approximation (RPA) energy, 
$U(\vec{q}_\perp)$, are respectively (after retrieving the spin index)
\begin{eqnarray}
U^{bind,\sigma_\alpha\sigma_\nu}
_{m_\alpha m_\mu;m_\nu m_\beta}(\vec{q}_\perp)&=&
\frac{1}{\Omega}\sum_{\vec{q}\,'}e^{i(q_xq_y'-q_yq_x')l_0^2}
V^{\sigma_\alpha\sigma_\nu}_{m_\alpha m_\mu;m_\nu m_\beta}(\vec{q}\,')
=-\frac{1}{\Omega}\sum_{\vec{q}\,'}\cos(q_xq_y'-q_yq_x')l_0^2)
V(\vec{q}\,')A^{\sigma_\alpha\sigma_\alpha}_{m_\alpha m_\mu}(-\vec{q}\,')
A^{\sigma_\nu\sigma_\nu}_{m_\nu m_\beta}(\vec{q}\,')
\label{U_bind}
\\
U^{\sigma_\alpha\sigma_\nu}_{m_\alpha m_\beta;m_\nu m_\mu}(\vec{q}_\perp)&=&
\frac{1}{2\pi l_0^2 L_z}\sum_{q_z'}
V^{\sigma_\alpha\sigma_\nu}_{m_\alpha m_\beta;m_\nu m_\mu}(\vec{q}_\perp,q_z')
=\frac{1}{2\pi l_0^2 L_z}\sum_{q_z'}
V(\vec{q},q_z')A^{\sigma_\alpha\sigma_\alpha}_{m_\alpha m_\beta}
(-\vec{q}_\perp,-q_z')
A^{\sigma_\nu\sigma_\nu}_{m_\nu m_\mu}(\vec{q}_\perp,q_z').
\label{U_RPA}
\end{eqnarray}
Note that the non-spin-flipping interaction, ($\sigma_\alpha=\sigma_\mu$,
$\sigma_\nu=\sigma_\beta$) for Eq. (\ref{U_bind}) and 
($\sigma_\alpha=\sigma_\beta$,$\sigma_\nu=\sigma_\mu$) for Eq. (\ref{U_RPA}),
is already incorporated above.
As mentioned in Sec. \ref{correlation_function},
instead of calculating the irreducible polarizability directly,
we derive the Dyson's equations of the interaction matrix element and 
a special vertex function in Eqs. (\ref{dynson_U2}) and (\ref{dynson_Lambda2}),
which can be used to obtain the collective mode energy and 
dielectric function.
Above derivation and results are independent of the
details of single particle wavefunctions or eigenenergies, and are valid 
for arbitrary quantum well confinement potential,
provided the relevant form function $A^{\sigma_i\sigma_j}
_{n_in_j}(\vec{q}\,)$ of 
Eq. (\ref{A_def}) is appropriately modified.
In considering the spin degree of freedom, only
non-spin-flip modes are included in Eq. (\ref{dynson_U2})
and Eq. (\ref{dynson_Lambda2}). The spin-flip modes do not 
include the bubble diagram due to the spin conservation implied by
the interaction shown 
in Eq. (\ref{dynson_U2}), and 
$W_{m_\alpha m_\beta;m_\nu m_\mu}(\vec{q}_\perp)$ of 
Eq. (\ref{W_def}) therefore becomes the same as the exciton binding energy,
$U^{bind}_{m_\alpha m_\mu;m_\nu m_\beta}(\vec{q}_\perp)$, for the same reason.
For convenience we will not distinguish these
two modes here until we get to the final results 
in the following section.

\subsection{Energy Dispersion of Magnetoplasmon Excitations: 
analytical expression} 
\label{pl_disperion}

Solving Eq. (\ref{dynson_Lambda2}) one can obtain the vertex function,
$\tilde{\Lambda}_{m_\mu m_\nu;n_2n_3}(\vec{q}_\perp,\omega)$, and substitute it
in Eq. (\ref{dynson_U2}) to get the full formula for
screened Coulomb interaction (using $\Lambda_{m_\mu m_\nu;n_2n_3}
(\omega,\vec{q}_\perp)
=U_{m_\mu m_\nu;n_2n_3}(\vec{q}_\perp)$ according to Eq. (\ref{Lambda})):
\begin{eqnarray}
&&\tilde{U}_{n_1n_4,n_2n_3}(\omega,\vec{q}_\perp)
\nonumber\\
&=&U_{n_1n_4,n_2n_3}(\vec{q}_\perp)
+\sum_{m_\alpha m_\beta}U_{n_1n_4;m_\beta m_\alpha}(\vec{q}_\perp)
D_{m_\alpha m_\beta}(\omega)\sum_{m_\mu m_\nu}
\left[\delta_{\alpha\mu}\delta_{\beta\nu}-{D}_{m_\alpha m_\beta}(\omega)
W_{m_\mu m_\nu;m_\beta m_\alpha}(\vec{q}_\perp)\right]^{-1}
\Lambda_{m_\mu m_\nu;n_2n_3}(\vec{q}_\perp,\omega)
\nonumber\\
&=&\sum_{m_\alpha m_\beta}U_{n_1n_4;m_\beta m_\alpha}(\vec{q}_\perp)
\left[\varepsilon^{-1}(\omega,\vec{q}_\perp)\right]
_{m_\alpha m_\beta;n_2n_3}, 
\label{U_screened}
\end{eqnarray}
where the dielectric function, $\epsilon(\omega,\vec{q}_\perp)$, 
is a matrix function,
\begin{eqnarray}
\left[\varepsilon(\omega,\vec{q}_\perp)\right]^{-1}
_{m_\alpha m_\beta;n_2n_3}&=&
\delta_{n_2m_\beta}\delta_{n_3m_\alpha}-
\sum_{m_\mu m_\nu}\left[
Y_{m_\alpha m_\beta,m_\mu m_\nu}(\omega,\vec{q}_\perp)\right]^{-1}
U_{m_\mu m_\nu;n_2n_3}(\vec{q}_\perp)
\label{matrix_dielectric},
\end{eqnarray}
and the "dispersion matrix", $Y$, is
\begin{eqnarray}
Y_{m_\alpha m_\beta,m_\mu m_\nu}(\omega,\vec{q}_\perp)\equiv
\left\{-\delta_{m_\alpha m_\mu}\delta_{m_\beta m_\nu}\left[
D_{m_\alpha m_\beta}(\omega)\right]^{-1}+
W_{m_\mu m_\nu;m_\beta m_\alpha}(\vec{q}_\perp)\right\}.
\label{Y_matrix}
\end{eqnarray}
The TDHF dynamical dielectric function appearing 
in Eq. (\ref{matrix_dielectric})
and Eq. (\ref{Y_matrix}) includes infinite series of 
both RPA bubble diagrams and the excitonic ladder diagrams. Theoretically 
given a finite matrix size by including relevant Landau levels
(i.e. by appropriately cutting off the infinite matrix equations give above),
one can numerically calculate
each element of the dielectric function and obtain the collective mode
dispersions by solving the standard collective mode equation,
$det\{\epsilon(\omega,\vec{q}_\perp)\}=0$.
However, it is easy to see from Eq. (\ref{matrix_dielectric}) that solving 
$\omega$ from $det\{\epsilon(\omega,\vec{q}_\perp)\}=0$ is the same as
solving $\omega$ from $det\{Y(\omega,\vec{q}_\perp)\}=0$, or more conveniently, 
the same as solving
the eigenvalue equation of $Y(0,\vec{q}_\perp)$ because 
$Y(\omega,\vec{q}_\perp)=Y(0,\vec{q}_\perp)+i\omega\cdot I$, where
$I$ is the identity matrix due to the special 
form of $D_{m_\alpha m_\beta}(\omega)$ in Eq. (\ref{D_w}).
Therefore focusing on the collective mode dispersion in this section, 
we will discuss the dispersion matrix
$Y(0,\vec{q}_\perp)$ below in more detail instead of the dielectric 
function itself, which is studied in the next section. We note that
this theoretical simplification of the equivalence between the 
simple static $Y$ function and the dynamical dielectric function 
in obtaining the collective mode dispersion has not earlier 
been appreciated in the literature.

\begin{figure}
 \vbox to 7cm {\vss\hbox to 5.5cm
 {\hss\
   {\includegraphics{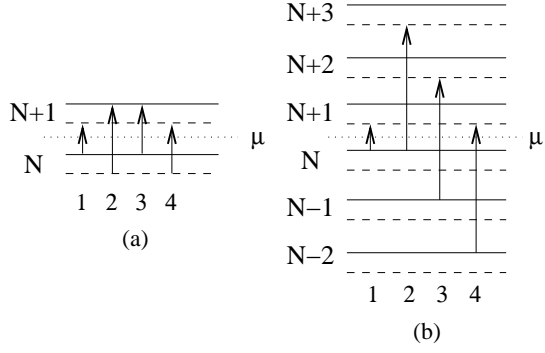}
   }
  \hss}
 }
\caption{
Energy level configuration for electron-hole pair
excitations. Solid(dashed) lines are for spin down(up) levels
with level index in the left hand side (the first orbital level
index is set to be zero), and the upward arrow represents an electron-hole
excitation (a magnetic exciton). (a) is for the two 
$2\times 2$ matrix representation of Eqs. (\ref{lowest_order_flip})
and (\ref{lowest_order_nonflip}): electron-hole pairs of numbers 1 and 2
are for $Y_\sigma$, and numbers 3 and 4 are for $Y_\rho$ respectively.
(b) shows the configuration for one spin-flip excitation
($\delta\sigma=+1$) including next higher order energy excitations, which
are beyond the TDHFA developed in the paper. 
(Note that in (b), the excitation from level 
$n$ to $n+2$ does not couple to pair
number 1 due to parity symmetry in a parabolic well.)
}
\label{energy_level_figure}
\end{figure}

According to Eq. (\ref{D_w}), the only valid matrix element of
Eq. (\ref{Y_matrix}) should be for the pair, $(m_\alpha,m_\beta)$,
of one electron in an empty level, $m_\alpha(m_\beta)$, and one hole in a
filled level, $m_\beta(m_\alpha)$. 
To the lowest order of $(e^2/\epsilon_0 l_0)/\omega_2$, only four
levels, $(N,\uparrow)$, $(N,\downarrow)$, $(N+1,\uparrow)$,
and $(N+1,\downarrow)$ are included 
(see Fig. \ref{energy_level_figure}(a)), where $N=\nu/2-1$ is the level
index of the highest filled level (note the first Landau
index has been taken to be zero). 
After separating spin-flip and non-spin-flip modes, 
one can obtain two $2\times 2$ $Y$ matrices for spin-flip ($\sigma$)
and non-spin-flip ($\rho$)
excitations respectively (after retrieving the spin index):
\begin{eqnarray}
Y_\sigma(\vec{q}_\perp)&=&
\left[\begin{array}{cc}
 \Delta E^{\downarrow\uparrow}_{NN+1}-U^{bind,\downarrow\uparrow}
_{NN,N+1N+1}(\vec{q}_\perp)& -U^{bind,\downarrow\downarrow}_{NN+1,NN+1}
(\vec{q}_\perp) \\
-U^{bind,\uparrow\uparrow}_{N+1N,N+1N}(\vec{q}_\perp)
&\Delta E^{\uparrow\downarrow}_{NN+1}-U^{bind,\uparrow\downarrow}
_{NN,N+1N+1}(\vec{q}_\perp) 
  \end{array}
\right],
\label{lowest_order_flip}
\\
Y_\rho(\vec{q}_\perp)&=&
\left[\begin{array}{cc}
 \Delta E^{\downarrow\downarrow}_{NN+1}-
U^{bind,\downarrow\downarrow}_{NN,N+1N+1}(\vec{q}_\perp)
+U^{\downarrow\downarrow}_{NN+1,N+1N}(\vec{q}_\perp)& 
U^{\downarrow\uparrow}_{NN+1,N+1N}(\vec{q}_\perp) \\
U^{\uparrow\downarrow}_{NN+1,N+1N}(\vec{q}_\perp) &
\Delta E^{\uparrow\uparrow}_{NN+1}
-U^{bind,\uparrow\uparrow}_{NN,N+1N+1}(\vec{q}_\perp)+
U^{\uparrow\uparrow}_{NN+1,N+1N}(\vec{q}_\perp)
  \end{array}
\right], 
\label{lowest_order_nonflip}
\end{eqnarray}
where $\Delta E^{\sigma_1\sigma_2}_{n_1n_2}=E^0_{n_2,\sigma_2}-
E^0_{n_1,\sigma_1}+\Sigma^{HF}_{n_2,\sigma_2}-\Sigma^{HF}_{n_1,\sigma_1}$ 
is the HF single particle energy difference. Note that the off-diagonal term
in $Y_\sigma(\vec{q}_\perp)$ is omitted in the paper
by Kallin and Halperin \cite{kallin} for a ZWW system, 
because they just considered the
leading order $1\times 1$ matrix representation of $Y_\sigma$. Using
$U^{bind,\downarrow\uparrow}_{N+1N,N+1N}(\vec{q}_\perp)=
\left[U^{bind,\downarrow\uparrow}_{NN+1,NN+1}(\vec{q}_\perp)\right]^\ast$,
we can obtain the dispersions of three spin collective modes and one charge
collective mode
accordingly by solving the determinantal equation for $Y$. 
(Note that for systems of even filling factor,
$\nu=2(N+1)$, the spin index can be neglected in the
self-energy, $\Sigma^{HF}_{n,\sigma}$, and interactions, $U$ and $U^{bind}$,
if the ground state is unpolarized and hence spin-symmetric):
\begin{eqnarray}
\omega_{\sigma_+}(\vec{q}_\perp)&=&
\omega_2+\Sigma^{HF}_{N+1}-\Sigma^{HF}_{N}
-U^{bind}_{NN,N+1N+1}(\vec{q}_\perp)-\sqrt{\omega_z^2+
\left| U^{bind}_{NN+1,NN+1}(\vec{q}_\perp)\right|^2/4}
\label{omega_+}
\\
\omega_{\sigma_-}(\vec{q}_\perp)&=&\omega_2
+\Sigma^{HF}_{N+1}-\Sigma^{HF}_{N}
-U^{bind}_{NN,N+1N+1}(\vec{q}_\perp)+\sqrt{\omega_z^2+
\left| U^{bind}_{NN+1,NN+1}(\vec{q}_\perp)\right|^2/4}
\label{omega_-}
\\
\omega_{\sigma_z}(\vec{q}_\perp)&=&\omega_2
+\Sigma^{HF}_{N+1}-\Sigma^{HF}_{N}
-U^{bind}_{NN,N+1N+1}(\vec{q}_\perp)
\label{omega_z}
\\
\omega_{\rho}(\vec{q}_\perp)&=&\omega_2
+\Sigma^{HF}_{N+1}-\Sigma^{HF}_{N}
-U^{bind}_{NN,N+1N+1}(\vec{q}_\perp)+2U_{NN+1,N+1N}(\vec{q}_\perp).
\label{omega_rho}
\end{eqnarray}

Equations (\ref{omega_+})-(\ref{omega_rho}) are our main analytical
results in this paper, 
which are the formal generalizations of the corresponding results 
in Ref. \cite{kallin} to a WWW in the tilted field,
except for the square-root term in Eqs. (\ref{omega_+})
and (\ref{omega_-}). This shows that the
off-diagonal term of $Y_\sigma(\vec{q}_\perp)$ is an 
exchange-interaction-induced level-repulsing 
effects for the two spin triplet modes, $\omega_{\sigma_\pm}$, and
effectively increases the Zeeman energy.
Note that the analytical derivation given above is not related
to any specific form of the single particle wavefunction which
enters only through the actual calculations of the various matrix elements
in Eqs. (\ref{omega_+})-(\ref{omega_rho}). The only constraint on the
wavefunctions is that they must be obtained in a conserving approximations.
There are several approximations we can use to obtain the single particle 
wavefunctions and eigenenergies. Here we will compare two of them:
one is the fully SCHF approximation as shown in Eq. 
(\ref{HF_equation_general}), and the other one is the first order
Hartree-Fock approximation, where electron Hartree and Fock 
potential are calculated by noninteracting electron wavefunctions, 
which are not renormalized by a self-consistent 
equation (see Fig. \ref{feynm_figure}(d)). It is shown later that such
a first order Hartree-Fock approximation does capture the most important 
contribution of the SCHFA, 
but is computationally much easier than the numerical results from
the SCHF equations. In fact, as mentioned before, 
the TDHFA in solving collective mode dispersion
is exact only to the leading order in the interaction. 
Therefore in some sense the 
{\it single} particle wavefunctions
and eigenenergies calculated in the full SCHFA are not guaranteed to give
better {\it collective} mode dispersion energies
than those calculated in the simple first order HF approximation, although
the former may very well be better in calculating the single
electron properties (such as the electron density profile or absorption spectra
\cite{dempsey}). In fact, we believe that in the spirit of our TDHFA
calculation for the collective mode dispersion, it is actually
{\it better} to use the first order HF wavefunctions and energies 
in the collective mode calculation in view of the excitations of 
the theory in the leading order Coulomb interaction.
The use of such first order HF wavefunctions and energies in the TDHFA
calculation of collective mode dispersions ensures that all quantities 
entering the theory are leading order in the Coulomb interaction \cite{rice65}.
Therefore it is instructive
to show the corresponding formula of the first order HF approximation
in our theory here and we will compare the two sets of numerical results 
(SCHF and first order HF) in the next section.
Defining the first order interaction
matrix element similar to Eqs. (\ref{bare_matrix}) and (\ref{A_def}) by
using the noninteracting wavefunctions in
Eq. (\ref{noninteracting-wavefunctions}), we have
\begin{eqnarray}
A^{(0),\sigma_i\sigma_j}
_{\vec{n}_i\vec{n}_j}(\vec{q}\,)&=&\int d\vec{r}\,e^{-i\vec{q}\cdot\vec{r}}
{\phi^0_{\vec{n}_i,-q_y/2,\sigma_i}}^\dagger(\vec{r}\,)
{\phi^0_{\vec{n}_j,q_y/2,\sigma_j}}(\vec{r}\,)
\nonumber\\
&=&\int dx\int dz\,e^{-iq_xx-iq_zz}
\Phi^0_{\vec{n}_i,\sigma_i}(x-q_yl_0^2/2,z)
\Phi^0_{\vec{n}_j,\sigma_j}(x+q_yl_0^2/2,z),
\label{A_def0}
\end{eqnarray}
whose analytical expression for a parabolic well could
be obtained by using the generalized Laguerre polynomial discussed in Appendix
\ref{A_function}. As a consequence, one can also obtain the analytical
expression corresponding to Eqs. (\ref{omega_-})-(\ref{omega_rho}) 
in the first order
Hartree-Fock approximation. For convenience,
we first define two new dimensionless quantities, 
$Q_1(\vec{q}\,)$ and $Q_2(\vec{q}\,)$,
as following:
\begin{eqnarray}
Q_1(\vec{q}\,)&=&\frac{\cos^2\theta(q_yl_0)^2+
(\cos\theta q_xl_0-\sin\theta q_zl_0)^2\lambda_1^2}{2\lambda_1}
\label{Q_1}
\\
Q_2(\vec{q}\,)&=&\frac{\sin^2\theta(q_yl_0)^2+
(\sin\theta q_xl_0+\cos\theta q_zl_0)^2\lambda_2^2}{2\lambda_2}
\label{Q_2}
.
\end{eqnarray}
The first order RPA (direct) energy then becomes (suppressing 
the spin index here):
\begin{eqnarray}
U^{(1)}_{NN+1,N+1N}(\vec{q}_\perp)&=&
\frac{1}{2\pi l_0^2 L_z}\sum_{q_z'}
V(\vec{q}_\perp,q_z')\left|A^{(0)}_{N+1N}(\vec{q}_\perp,q_z')\right|^2
\nonumber\\
&=&\frac{1}{2\pi l_0^2 L_z(n+1)}\sum_{q_z'}
V(\vec{q},q_z')
\exp\left[-Q_1(\vec{q}_\perp,q'_z)\right]
\exp\left[-Q_2(\vec{q}_\perp,q'_z)\right]
Q_2(\vec{q}_\perp,q'_z)
\left|
L_N^1\left(Q_2(\vec{q}_\perp,q'_z)\right)
\right|^2,
\label{U_explicit}
\end{eqnarray}
where $L_n^m(x)$ is the generalized Laguerre polynomial 
and the first order exciton binding (exchange) energy is
\begin{eqnarray}
U^{(1),bind}_{NN,N+1N+1}(\vec{q}_\perp)&=&
\frac{1}{\Omega}\sum_{\vec{q}\,'}\cos(q_xq_y'-q_yq_x')l_0^2)
V(\vec{q}\,')A^{(0)}_{NN}(-\vec{q}\,')A^{(0)}_{N+1N+1}(\vec{q}\,')
\nonumber\\
&=&\frac{1}{\Omega}\sum_{\vec{q}\,'}\cos(q_xq_y'-q_yq_x')l_0^2)
V(\vec{q}\,')\exp\left[-Q_1(\vec{q}\,'_\perp)\right]
\exp\left[-Q_2(\vec{q}\,'_\perp)\right]
L_{N}^0\left(Q_2(\vec{q}\,'_\perp)\right)
L_{N+1}^0\left(Q_2(\vec{q}\,'_\perp)\right)
\label{U'_explicit}
\end{eqnarray}

For the first order HF self-energy, it is more convenient and instructive to
show the self-energy difference between levels
$N$ and $N+1$ individually for the direct or the Hartree term ($\Sigma^H$)
and the exchange or the Fock term ($\Sigma^F$):
\begin{eqnarray}
\Sigma^{(1),H}_{N+1}-\Sigma^{(1),H}_N&=&
\frac{2}{2\pi l_0^2L_z}\sum_{q_z}\sum_{l=0}^NV(q_z)A^{(0)}_{{l}{l}}
(q_z)\left[A^{(0)}_{N+1N+1}(q_z)-A^{(0)}_{NN}(q_z)\right]
\nonumber\\
&=&\frac{2}{2\pi l_0^2L_z}\sum_{q_z}V(q_z)
\exp\left[-Q_1(q'_z)\right]
\exp\left[-Q_2(q'_z)\right]
\left[L_{N+1}^0\left(Q_2(q'_z)\right)
-L_N^0\left(Q_2(q'_z)\right)\right]
\sum_{l=0}^N L_l^0\left(Q_2(q'_z)\right),
\label{Hartree_self_explicit}
\end{eqnarray}
\begin{eqnarray} 
&&\Sigma^{(1),F}_{N+1}-\Sigma^{(1),F}_N=
\frac{-1}{\Omega}\sum_{\vec{q}}
V(\vec{q}\,)\left[\sum_{l=0}^N
\,|A^{(0)}_{l,N+1}(\vec{q}\,)|^2 - \sum_{l=0}^N
\,|A^{(0)}_{lN}(\vec{q}\,)|^2\right]
\nonumber\\
&=&\frac{-1}{\Omega}\sum_{\vec{q}}
V(\vec{q}\,)\exp\left[-Q_1(\vec{q}\,)\right]
\exp\left[-Q_2(\vec{q}\,)\right]
\frac{1}{N!}\sum_{l=0}^N(Q_2(\vec{q}\,))^{N-l} l!
\left[\frac{1}{N+1}Q_2(\vec{q}\,)
\left|L^{N+1-l}_l(Q_2(\vec{q}\,))\right|^2
-\left|L^{N-l}_l(Q_2(\vec{q}\,))\right|^2\right].
\label{Fock_self_explicit}
\end{eqnarray}
It is easy to prove that
\begin{eqnarray}
\Sigma^{(1),H}_{N+1}-\Sigma^{(1),H}_N&=&
-2U^{(1)}_{NN+1,N+1N}(\vec{0}) \nonumber\\
\Sigma^{(1),F}_{N+1}-\Sigma^{(1),F}_N&=&U^{(1),bind}_{NN,N+1N+1}(\vec{0})
\label{kohn_theorem}
\end{eqnarray}
by using the following two identities for the generalized Laguerre polynomials:
\begin{eqnarray}
xL_{n}^{m+1}(x)&=&(n+m+1)L_n^m(x)-(n+1)L_{n+1}^m(x) \nonumber\\
\sum_{l=0}^nL_l^m(x)&=&L_n^{m+1}(x).
\end{eqnarray}
Eq. (\ref{kohn_theorem}) shows that in the long wavelength limit, 
the charge density collective mode has the same energy as its noninteracting
result (as it must),
\begin{eqnarray}
\omega^{(1)}_\rho(\vec{q}_\perp\rightarrow 0)&=&
E^0_{0,N+1,\downarrow}-E^0_{0,N,\uparrow}=\omega_2,
\label{general_kohn}
\end{eqnarray}
which reflects
the generalized Kohn's theorem \cite{brey89}. This shows
that the time-dependent Hartree-Fock approximation we apply in this paper
is a current-conserving approximation to the leading order single electron
wavefunctions and eigenenergies. 
From the numerical calculation presented in the next section, 
such a generalized Kohn's theorem, $\omega_\rho(\vec{q}_\perp
\rightarrow 0)=\omega_2$,
is true also for Eq. (\ref{omega_rho}), where the electron 
wavefunction is calculated self-consistently through Eq. 
(\ref{HF_equation_general}). 
However, one should note that if one includes the
larger matrix size in Eq. (\ref{Y_matrix}) to go beyond the lowest
order in $(e^2/\epsilon_0 l_0)/\omega_2$ (see Fig. 
\ref{energy_level_figure}(b)), 
there is no such exact
cancelation, since some more diagrams (higher order in the interaction)
should be included in Fig.
\ref{feynm_figure} in order to obtain the
current-conserving theory for collective modes in higher order calculations.

\subsection{Energy Dispersions of Magnetoplasmon Excitations: 
numerical results}
\label{plasmon_numerical}

\begin{figure}
 \vbox to 10cm {\vss\hbox to 5.5cm
 {\hss\
    {\includegraphics{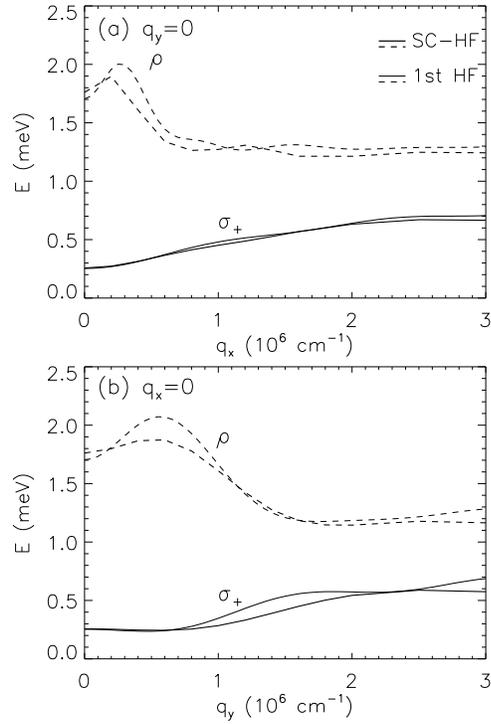}
  }
  \hss}
  }
\caption{
Magnetoplasmon dispersions for $\nu=6$ and $B_\|$ (along $x$ direction)
is 11 Tesla, calculated
from Eqs. (\ref{omega_-}) and (\ref{omega_rho}).
(a) and (b) are for momentum along $x$ and $y$ directions respectively.
Thick(thin) lines are for wavefunctions calculated from self-consistent
Hartree-Fock and from first order Hartree-Fock approximations respectively.
}
\label{unscrn_2l_cs_n6_w7}
\end{figure}

The two $2\times 2$ matrices shown in Eqs. (\ref{lowest_order_flip})
and (\ref{lowest_order_nonflip}) 
give different magnetoplasmon excitation branches:
three triplet spin-density-excitations (denoted by 
$\omega_{\sigma_\pm}$ and $\omega_z$), and one
singlet charge-density-excitation (denoted by $\omega_\rho$).
In Fig. \ref{unscrn_2l_cs_n6_w7} we show the calculated dispersion
energies of the charge mode, 
$\omega_{\rho}(\vec{q}_\perp)$, and the lowest energy triplet
spin mode, $\omega_{\sigma_+}(\vec{q}_\perp)$, for a typical
parallel magnetic field, $B_\|=11$ Tesla at filling factor $\nu=6$
and other system parameters chosen to correspond to
the experimental sample \cite{pan01}.
The most important feature in the spectra is that
there is an energy minimum ("magneto-roton") at a finite wavevector,
$q_y^\ast\sim l_2^{-1}$, in the spin mode dispersion along
$y$ direction (perpendicular to the in-plane magnetic field which is
along the $x$ axis), while no such finite wavevector
minimum exists along the $x$ direction. 
Comparing with the zero width 2D
results (without any in-plane field) obtained in \cite{kallin}, 
where a roton-minimum is found
in the spin mode along {\it both} directions, one finds that the finite 
width reduces the 
electron-hole binding energy (Eq. (\ref{U_bind})), which is the origin
of the roton-minimum in the magnetic exciton picture, 
along the direction of the in-plane
magnetic field. (Note that for a ZWW system, the in-plane magnetic field
does not change the electron orbital wavefunctions and it simply
increases the Zeeman energy only, which is proportional to the total 
magnetic field) . From the energetic point of view,
therefore, this "softening" associated with the development of
the roton minimum (transverse to the in-plane field direction)
implies that the ground state of such a quantum Hall system
has a tendency to to make a transition from a uniform, unpolarized state 
to a spin density wave state
with broken translational and spin symmetries, particularly 
if this roton-minimum
reaches zero energy in some situations. In our calculation,
the minimum energy of the spin mode ($\omega_{\sigma_+}$)
goes to zero energy at $B_\|=12.5$ Tesla. However, this value
is close to, but slightly larger than the critical in-plane magnetic field,
$B_\|^\ast=11.1$ meV, where the ground state makes the first order spin
polarization transition from a paramagnetic ($\nu_\uparrow=\nu_\downarrow$)
to the spin-polarized state ($\nu_\downarrow=\nu_\uparrow-2$)
(see Table 
\ref{critical_B_table}) in the Hartree-Fock approximation. 
Therefore, within our HF approximation the roton-minimum of the
spin mode dispersion does not actually go to zero energy before the whole
system undergoes a first order phase transition to a polarized ground state.
Calculating the collective mode energies for a {\it polarized} ground state
after level crossing, we find that this roton minimum energy
does not vanish, and in fact, may even increase in magnitude.
Therefore we do not observe 
a true mode softening in the spin density excitation in 
the present Hartree-Fock approximation although we see a clear tendency toward
such a possibility within our HF theory.
It is certainly possible that a more sophisticated 
approximation going beyond the HF approximation would produce
such mode softening (see the discussion in
the following sections). Note that the charge collective mode energy in
the long wavelength limit is exactly the same as the noninteracting 
energy separation, $\omega_2=1.72$ meV, in Fig. \ref{unscrn_2l_cs_n6_w7}, for
results calculated in both the first order HF approximation and
the SCHFA, within a 3\% numerical error.
As should be obvious from our results, there is no qualitative 
difference whatsoever between the results in these two 
approximations, which is not unexpected. Therefore, from now on, we will
only show results obtained in the first order HF approximation,  
not only because of its computational simplicity (saving considerable
time in numerical calculations), but also because, as mentioned in Section
\ref{pl_disperion}, we believe that the leading order HF calculation 
is really more consistent with our TDHFA theory for the collective modes.

\begin{figure}
 \vbox to 10cm {\vss\hbox to 5.5cm
 {\hss\
    {\includegraphics{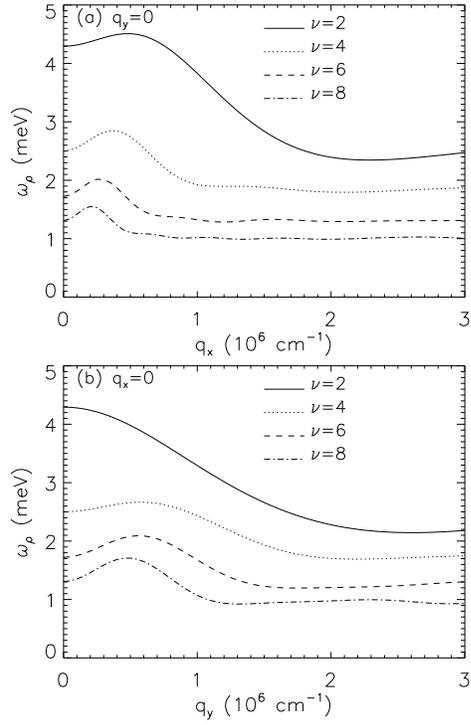}
  }
  \hss}
  }
\caption{
Charge mode dispersion, $\omega_\rho(\vec{q}_\perp)$,
of magnetoplasmon excitations of the same system as used in
Fig. \ref{unscrn_2l_cs_n6_w7} but for different filling factors,
$\nu=2$, 4, 6 and 8, for comparison. 
}
\label{unscrn_2l_c_w7}
\end{figure}

\begin{figure}
 \vbox to 10cm {\vss\hbox to 5.5cm
 {\hss\
    {\includegraphics{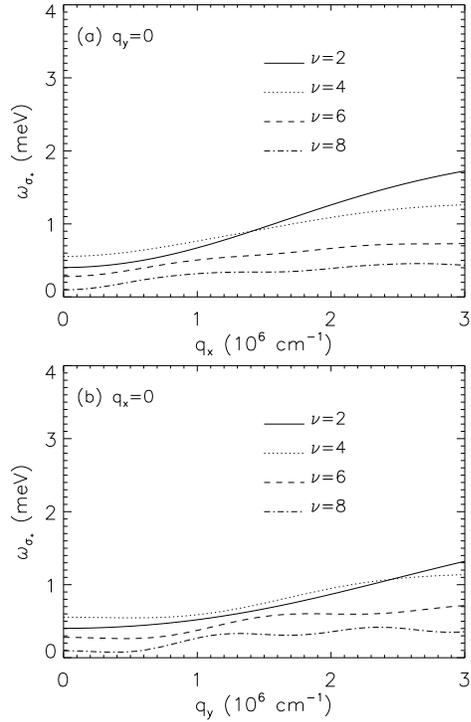}
  }
  \hss}
  }
\caption{
Same as Fig. \ref{unscrn_2l_c_w7}, but for spin mode, $\omega_{\sigma_+}
(\vec{q}_\perp)$, dispersion.
}
\label{unscrn_2l_s_w7}
\end{figure}

In Figs. \ref{unscrn_2l_c_w7} and \ref{unscrn_2l_s_w7} we show respectively
the charge and spin mode dispersions for $\nu=2$, 4, 6, and 8 system 
by changing $B_\perp$ (total electron density is fixed) 
with all other parameters the 
same as in Fig. \ref{unscrn_2l_cs_n6_w7}.
The RPA peak is relatively weaker in stronger perpendicular
magnetic field (smaller $\nu$), while it is more pronounced when
more Landau levels are occupied (larger $\nu$).
On the other hand, the energy difference
between the long wavelength limit (which is just the noninteracting energy 
gap, $\omega_2$, according to Eq. (\ref{kohn_theorem}))
and the roton minimum of the charge mode excitation is larger for
smaller $\nu$ (stronger $B_\perp$) system. 
This indicates that the multiple 
absorption peaks observed in the polarized inelastic light 
scattering experiment \cite{pinczuk88} 
should be separated more widely for smaller
$\nu$ (stronger $B_\perp$).
For the spin mode excitations shown in Fig. \ref{unscrn_2l_s_w7}, 
the results for different filling factors are quite similar,
except for their different $q^\ast_y$ (i.e. the position of the 
magneto-roton minima) due to different 
perpendicular magnetic field values.

\begin{figure}
 \vbox to 10cm {\vss\hbox to 5.5cm
 {\hss\
    {\includegraphics{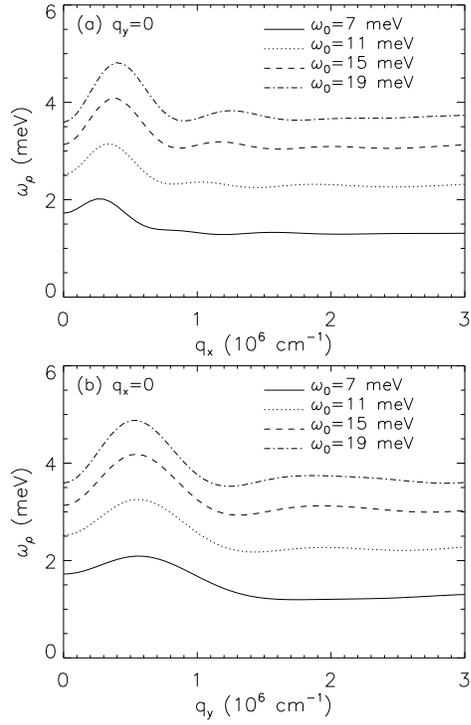}
  }
  \hss}
  }
\caption{
Charge mode dispersion
of magnetoplasmon excitation of the same system as used in
Fig. \ref{unscrn_2l_cs_n6_w7} but with different
confinement energy, $\omega_0$,
at filling factor, $\nu=6$.
Zero field well widths are about 260, 200, 175, and 155\AA,
corresponding to $\omega_0=7$, 11, 15, and 19 meV respectively.
The parallel magnetic field is 11 Tesla for all results.
}
\label{unscrn_2l_c_n6}
\end{figure}

\begin{figure}
 \vbox to 10cm {\vss\hbox to 5.5cm
 {\hss\
    {\includegraphics{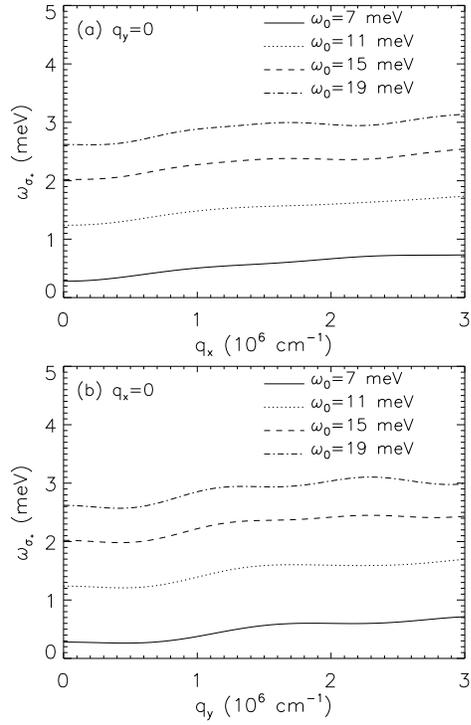}
  }
  \hss}
  }
\caption{
Same as Fig. \ref{unscrn_2l_c_n6}, but for spin mode dispersion.
}
\label{unscrn_2l_s_n6}
\end{figure}

In Figs. \ref{unscrn_2l_c_n6} and \ref{unscrn_2l_s_n6} we show respectively
the charge and spin mode dispersions for $\nu=6$ system but with different 
confinement energy, $\omega_0$, as indicated in the figures.
Larger confinement energies indicate smaller well widths 
in $z$ direction. Therefore we have a continuous "transition" from
3D to 2D by increasing $\omega_0$ at a fixed density.
This transition is observed clearly in Figs.
\ref{unscrn_2l_c_n6} and \ref{unscrn_2l_s_n6} where
the spectra in $x$ and $y$ directions become very similar
for higher values of $\omega_0$, reproducing the zero width 
(strictly 2D) results \cite{kallin}.
On the other hand, the roton minimum energy of the
$\omega_{\sigma_+}$ mode decreases for weaker confinement potential
(larger effective well width), showing more of a tendency to have a 
spin-density-wave instability in a wider well.
Another important feature can be seen in the charge mode dispersion.
When the confinement potential is weak
(e.g. $\omega_0=7$ meV), the energy of the roton-minimum
is smaller than the mode energy in the long wavelength 
limit ($\vec{q}_\perp=0$).
But the roton energy becomes larger than the long wavelength mode energy
when the confinement potential is increased to $\omega_0=19$ meV, reproducing
the results of the pure 2D system \cite{kallin}, where the roton minimum
is typically at a higher energy than the long wavelength mode energy.
Therefore the finite width effect also enhances
the tendency of a charge density wave instability against the ground state.

\begin{figure}
 \vbox to 10cm {\vss\hbox to 5.5cm
 {\hss\
    {\includegraphics{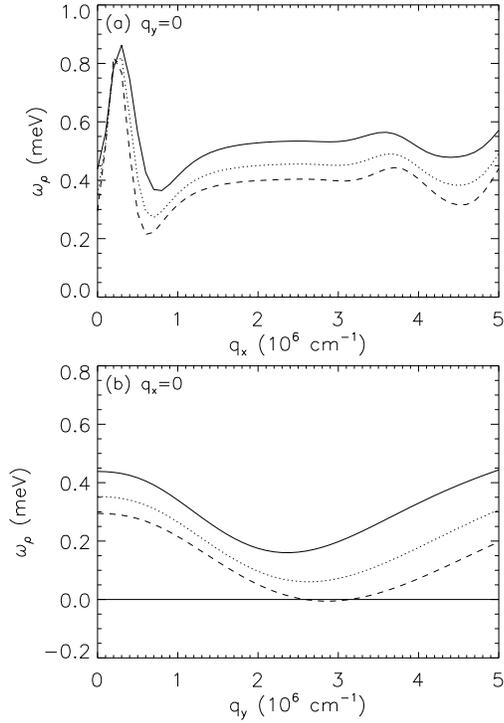}
  }
  \hss}
  }
\caption{
Charge mode magnetoplasmon dispersions for $\nu=1$, $B_\perp$=3 Tesla, and 
$\omega_0=3$ meV, calculated in the TDHFA. 
Solid, dotted, and dashed lines are for $B_\|=20$, 25, and 30 Tesla
respectively, showing a charge mode softening 
in $y$ direction, perpendicular to
the $B_\|$ direction. 
(a) and (b) are for wavevectors along $x$ and $y$ directions respectively.
}
\label{pl_nu1}
\end{figure}

In Fig. \ref{pl_nu1} we show a typical singlet charge density magnetoplasmon
mode ($\omega_\rho$) dispersion of $\nu=1$ as an example of odd 
filling factors in TDHFA. For $\nu=1$, there will be no first order 
phase transition by Landau level crossing in {\it any} strength of in-plane
magnetic field. When $B_\|$ is more than 30 Tesla, we find a charge 
density wave instability at a finite wavevector perpendicular to the
in-plane magnetic field. More detailed Hartree-Fock analysis shows that
\cite{unpublished} this CDW is a kind of isospin skyrmion stripe,
which has a charge density modulation in the $x-y$ plane as discussed
in Section \ref{discussion}. 

As a final remark,
we note that Eq. (\ref{matrix_dielectric}) and Eq. (\ref{Y_matrix})
are based on TDHFA, which is exact only to the lowest order in the ratio of 
interaction energy to noninteracting energy gap 
($(e^2/\epsilon_0 l_0)/\omega_2$). Therefore it is a priori not clear
if this leading-order many-body approximation can be used to study
the mode softening phenomena near level crossing, where the interaction
energy is necessarily comparable to (or stronger than) the 
noninteracting level separation since the noninteracting levels becomes
degenerate at the critical point.
However, to the best of our knowledge, no other systematic reliable
technique is available to calculate the collective mode energy and
such a mode softening behavior was earlier successfully treated within the
TDHFA in the context of
the second order phase transition related to
the canted antiferromagnetic phase of a double layer system
in the presence of interlayer tunneling and Zeeman splitting 
\cite{canted_phase}.
Therefore we believe our results should be qualitatively valid
in the level crossing regime. We do not, however, exclude the 
possibility that correctly
including higher order interaction effects may very well 
reduce the roton-minimum energy
zero at a finite wavevector before the system undergoes 
a first order phase transition to a polarized ground state.
We do not know how to go beyond the TDHFA in a systematic current-conserving
manner but speculate that such a calculation may very well give rise 
to a finite wavevector softening of the magneto-roton producing a 
quantum phase transition to the symmetry-broken phase. Our 
speculation is partly based on our finding that TDHFA 
actually predicts such a transition at $B_\|=B_\|^\ast$ 
which happens to be sightly larger than the critical field 
for the first order transition.

\section{Screening Effects}
\label{screening}

In the TDHFA shown in the previous sections, electron-electron
interaction is the the bare Coulomb interaction without taking into account
screening effects from the electron-hole fluctuations in the Landau levels.
In this section, we will incorporate screening effects
in our magnetoplasmon calculations.
Actually, in Eq. (\ref{matrix_dielectric}) and
(\ref{Y_matrix}), a complete formula for
the dielectric function in TDHFA has been given, but this formula is
in general too complicated
to be widely used in an integer quantum Hall system.
In this section we will 
derive some convenient formulae for the dielectric function, $\epsilon
(\vec{q},\omega)$, in different reasonable limits. 
Including such screening effects in the bare
Coulomb interaction one may study the magnetoplasmon excitations
beyond the time-dependent Hartree-Fock approximation, where the
interaction used in the Green's function and vertex function is 
the unscreened one (see Figs. 
\ref{feynm_figure} and \ref{other_diagram}(a),(b)). 
For convenience of discussion, 
we will first show the results for a zero width well,
the system most theoretical researchers consider in the
literature, and then the screening results for the 
WWW system of interest to us.
It is shown below that for a ZWW, we can obtain a "scalar"
(not matrix) dielectric function including both RPA and ladder
diagrams shown in Fig. \ref{feynm_figure}(c) in TDHFA, 
i.e. the screening
effect in a ZWW is independent of the level index within
TDHFA. This result is valid beyond the pure RPA result proposed before
in Ref. \cite{aleiner95} and
should apply even at low density, where only a few 
Landau levels are occupied, because of
the inclusion of the ladder diagrams (left out 
in Ref. \cite{aleiner95}). For a WWW,
instead of using the complete result shown in Eq. (\ref{matrix_dielectric}) and 
(\ref{Y_matrix}), we will derive a conventional formula
in the strong parallel magnetic field region, which in some sense is effectively
similar to a ZWW system as mentioned in Sec. \ref{noninteracting}.
An analytical expression for the dielectric function 
can be obtained when only the RPA screening is
considered (neglecting ladder diagrams) and is a good approximation
for high density systems. Note that
these general formula of screening effects
could be used to study other interaction-induced electronic
properties of quantum Hall systems \cite{stanescu00,lopatnikova01},
and are therefore of broad general interest in quantum Hall 
problems transcending the specific applications we are 
dealing with in this paper.

\subsection{Screening in a Zero Width Well}
\label{screening_zero_width}

For a strictly 2D ZWW one can neglect the $z$ degree of freedom completely,
and therefore the interaction matrix element of
Eq. (\ref{bare_matrix}) and Eq. (\ref{U_def}) can be simplified to the
product of Coulomb interaction and the function $A^{2D}_{nm}(q)$:
\begin{eqnarray}
U^{2D}_{n_1n_4;n_2n_3}(\vec{q}_\perp)=V^{2D}(\vec{q}_\perp)
A^{2D}_{n_1n_4}(-\vec{q}_\perp)A^{2D}_{n_2n_3}(\vec{q}_\perp)
=V^{2D}(q)A^{2D}_{n_1n_4}(q)A^{2D}_{n_2n_3}(q),
\label{UVA}
\end{eqnarray}
where ${V}^{2D}(\vec{q}_\perp)\equiv (2\pi l_0^2 L_z)^{-1}\int dq_z\,
V(\vec{q}_\perp,q_z)$ is the two dimensional Coulomb interaction and
$A^{2D}_{n_in_j}(\vec{q}_\perp)=A^{2D}_{n_in_j}({q})$ is obtained
by using the standard Landau level 2D single particle 
wavefunction in an integral
similar to Eq. (\ref{A_def}). Its explicit 
formula can be obtained by taking the zero width limit 
($\omega_0\rightarrow \infty$) of the function
$A^{(0)}_{n_in_j}(\vec{q}\,)$ in Appendix \ref{A_function}.
Note that in such a pure 2D system, electron 
wavefunctions obtained by SCHFA are 
exactly the same as the noninteracting wavefunctions, so that
the results in the SCHFA and in the first order HFA are the same in this case.
We use the superscript, "2D", to denote pure two-dimensional
quantities in zero well width limit, and replace 
$\vec{q}_\perp$ by its absolute value, $q$, in Eq. (\ref{UVA})
due to the rotational symmetry in the $x-y$ plane in the 2D limit.

Applying Eq. (\ref{UVA}) to Eq. (\ref{U_screened}) and Eq. 
(\ref{matrix_dielectric}), one obtains immediately
\begin{eqnarray}
\tilde{U}^{2D}_{n_1n_4,n_2n_3}(\omega,q)
&=&U^{2D}_{n_1n_4,n_2n_3}(\omega,q)
\left(1-V^{2D}(q)\sum_{m_\alpha m_\beta}A^{2D}_{m_\beta m_\alpha}(q)
\sum_{m_\mu m_\nu}\left[
Y^{2D}_{m_\alpha m_\beta,m_\mu m_\nu}(\omega,q)\right]^{-1}
A^{2D}_{m_\mu m_\nu}(q) \right),
\label{U_zerowidth}
\end{eqnarray}
where the matrix $Y^{2D}(\omega,q)$ is of the same form as $Y(\omega,q)$
in Eq. (\ref{Y_matrix}), but now with 
two-dimensional interaction matrix elements. 
The dielectric function for a ZWW system is therefore
a scalar function and independent of the level index:
\begin{eqnarray}
\epsilon^{2D}(\omega,q)=
\left(1-V^{2D}(q)\sum_{m_\alpha m_\beta}A^{2D}_{m_\beta m_\alpha}(q)
\sum_{m_\mu m_\nu}\left[
Y^{2D}_{m_\alpha m_\beta,m_\mu m_\nu}(\omega,q)\right]^{-1}
A^{2D}_{m_\mu m_\nu}(q) \right)^{-1},
\label{dielectric_zerowidth}
\end{eqnarray}
and the corresponding irreducible 
polarizability, $\Pi^{2D}_{irr}(\omega,q)$ can be easily
obtained by using
$\epsilon^{2D}(\omega,q)=1-V^{2D}(q)\Pi^{2D}_{irr}(\omega,q)$. 
Note that when the spin degree of freedom is considered in
Eq. (\ref{dielectric_zerowidth}), all
electron-hole pair fluctuations involved in $\epsilon^{2D}(\omega,q)$
should be non-spin-flip pairs because Coulomb
interaction does not flip electron spin.

It is instructive to study the forms for the 
dielectric function in some special limits. First,
in the low frequency region, where only fluctuations like
$(N,\downarrow)\to (N+1,\downarrow)$ and 
$(N,\uparrow)\to (N+1,\uparrow)$ are relevant,
we can use the $2\times 2$ matrix of $Y_\rho(\omega,q)$ 
in Eq. (\ref{lowest_order_nonflip})
to express the dielectric function in the lowest order 
of $(e^2/\epsilon_0 l_0)/\omega_2$:
\begin{eqnarray}
\epsilon^{2D}(\omega\to 0,q)
&\sim&\left(
1-V^{2D}(q)\left|A^{2D}_{N,N+1}(q)\right|^2
\sum_{i,j}\left[
Y^{2D}_{i,j}(\omega,q)\right]^{-1}\right)^{-1} 
\nonumber\\
&=&\left(1-\frac{2 V^{2D}(q)\left|A^{2D}_{N,N+1}(q)\right|^2}
{\Delta E^{\downarrow\downarrow,2D}_{NN+1}-
U^{bind,2D}_{NN,N+1N+1}(q)+2U^{2D}_{NN+1,N+1N}(q)
+i\omega}\right)^{-1}
\nonumber\\
&=&\frac{\omega^{2D}_{\rho}(q)+i\omega}{\omega^{2D}_{\sigma_z}(q)+i\omega},
\label{static_zerowidth}
\end{eqnarray}
where we have used the fact that $\Delta E^{\downarrow\downarrow,2D}_{nn+1}=
\Delta E^{\uparrow\uparrow,2D}_{nn+1}$ for systems with 
even filling factors in the unpolarized ground state. 

Another good approximation for the dielectric function 
of Eq. (\ref{dielectric_zerowidth}) can be obtained in the high  
density limit, where it is well-known that the contribution of 
RPA diagrams dominates that of ladder diagrams in the correlation energy
\cite{fetter}.
Starting from Eq. (\ref{dynson_U2}) and (\ref{dynson_Lambda2}) and 
using iterations with Eq. (\ref{UVA}) to represent 
$W^{2D}_{m_\alpha m_\beta,m_\nu m_\mu}(q)$, which is now the same
as $U^{2D}_{m_\alpha m_\beta,m_\nu m_\mu}(q)$, we have
\begin{eqnarray}
\tilde{U}^{2D}_{n_1n_4,n_2n_3}(q,\omega)
&\sim&U^{2D}_{n_1n_4,n_2n_3}({q})
+U^{2D}_{n_1n_4,n_2n_3}({q})
\sum_{m_\alpha m_\beta}U^{2D}_{m_\alpha m_\beta;m_\beta m_\alpha}({q})
D^{2D}_{m_\alpha m_\beta}(\omega) 
\nonumber\\
&&+U^{2D}_{n_1n_4,n_2n_3}({q})\sum_{m_\alpha m_\beta}
U^{2D}_{m_\alpha m_\beta;m_\beta m_\alpha}({q})
D^{2D}_{m_\alpha m_\beta}(\omega)\cdot \sum_{m_\mu m_\nu}
U^{2D}_{m_\mu m_\nu;m_\nu m_\mu}({q})
D^{2D}_{m_\mu m_\nu}(\omega)+\cdots
\nonumber\\
&=&U^{2D}_{n_1n_4,n_2n_3}({q})\left[1-
\sum_{m_\alpha m_\beta}U^{2D}_{m_\alpha m_\beta;m_\beta m_\alpha}({q})
D^{2D}_{m_\alpha m_\beta}(\omega)\right]^{-1}.
\end{eqnarray}
After retrieving the spin index, we obtain
\begin{eqnarray}
\epsilon^{2D}_{RPA}(\omega,q)&=&1-
\sum_{m_\alpha m_\beta}U^{2D}_{m_\alpha m_\beta;m_\beta m_\alpha}({q})
D^{2D}_{m_\alpha m_\beta}(\omega)
\nonumber\\
&=&1+V^{2D}({q})
\sum_{\sigma}\sum_{m_=0}^{N_\sigma}\sum_{n=N_\sigma+1}^{\infty}
\frac{2(E_{n,\sigma}-E_{m,\sigma})\,\omega_\perp}
{(E_{n,\sigma}-E_{m,\sigma})^2+\omega^2}
\frac{m !}{n !}\left(\frac{q^2l_0^2}{2}\right)^{n-m}
\,e^{-q^2l_0^2/2}\left[L_{m}^{n-m}
\left(\frac{q^2l_0^2}{2}\right)\right]^2,
\label{pi_aleiner}
\end{eqnarray}
which is the same as the result in Ref. \cite{aleiner95} 
(using the identity: $L_{n+m}^{-m}(x)=(-1)^m (n!/(n+m)!) x^m L_n^m(x)$),
if we neglect the self-energy correction in the 
single particle energy, $E_{n_\sigma,\sigma}$ (so that
$E_{n_\sigma,\sigma}=E^0_{n_\sigma,\sigma}=
(n_\sigma+1/2)\omega_\perp-\sigma\omega_z$).
Note that the RPA result shown in Eq. (\ref{pi_aleiner}) 
includes the dressed single particle Green's function via 
the Fock self-energy correlation (the Hartree term is cancelled), but
it sums over all empty and filled levels and is therefore actually beyond
the validity range of TDHFA which neglects multi-exciton 
effects. Both Eq. (\ref{dielectric_zerowidth}) and Eq.
(\ref{pi_aleiner}) above are independent of the parallel (in-plane) 
magnetic field in the strict 2D limit, since
the parallel magnetic field only affects 
the Zeeman energy in the strict 2D limit (and not any 
aspects of the orbital motion). However, the parallel field does,
as expected, affect the dielectric function in a
finite width well as shown below.

\subsection{Screening in a Wide Quantum Well}
\label{screening_finite_width}

For a WWW (specifically a parabolic WWW for our calculations), 
Eqs. (\ref{matrix_dielectric})
and (\ref{Y_matrix}) show that the dielectric function
is a matrix function strongly dependent on the level 
index of the interaction matrix 
element. In general, these expressions are not convenient
for applications in different physical problems, and therefore 
we have to look for a good approximation for Eq. (\ref{matrix_dielectric}).
First we could
get a good low frequency approximation for the dielectric function
by truncating the matrix size of Eq. (\ref{matrix_dielectric})
into $2\times 2$ and applying  
$Y_\rho$ of Eq. (\ref{lowest_order_nonflip}), i.e. considering electron-hole
fluctuations only between the two nearest levels about the Fermi level.
The result is similar to Eq. (\ref{static_zerowidth}):
\begin{eqnarray}
\epsilon_{N,N+1}(\omega\to 0,\vec{q}_\perp)
&\sim&\frac{\omega_{\rho}(\vec{q}_\perp)+i\omega}
{\omega_{\sigma_z}(\vec{q}_\perp)+i\omega},
\label{static_finitewidth}
\end{eqnarray} 
where $\omega_{\sigma_z}(\vec{q}_\perp)$
and $\omega_{\rho}(\vec{q}_\perp)$ are given by Eq. (\ref{omega_z})
and Eq. (\ref{omega_rho}) respectively. The difference between 
Eq. (\ref{static_zerowidth}) for a ZWW system and 
Eq. (\ref{static_finitewidth}) for a WWW system is that the 
former can be used for interaction between electrons in any Landau levels,
while the later is correct only for electrons interacting between
$(N,\uparrow(\downarrow))$ and $(N+1,\uparrow(\downarrow))$ levels
in the low energy region of an unpolarized ground state. 
When considering higher energy excitation,
say electrons from $N-2$ level to level $N+1$, a larger matrix 
representation for the $Y_\rho$ matrix has to be used to get a 
self-consistent result, but it may exceed the validity region of 
TDHFA. It is instructive to check the asymptotic approximation of
Eq. (\ref{static_finitewidth}) in the static long wavelength limit by
using Eqs. (\ref{omega_z}), (\ref{omega_rho}), and (\ref{kohn_theorem}):
\begin{eqnarray}
\epsilon_{N,N+1}(0,\vec{q}_\perp\rightarrow 0)
&\rightarrow&\frac
{\omega_2+\Sigma^{HF}_{N+1}-\Sigma^{HF}_{N}
-U^{bind}_{NN,N+1N+1}(\vec{q}_\perp\rightarrow 0)+
2U_{NN+1,N+1N}(\vec{q}_\perp\rightarrow 0)} 
{\omega_2+\Sigma^{HF}_{N+1}-\Sigma^{HF}_{N}
-U^{bind}_{NN,N+1N+1}(\vec{q}_\perp\rightarrow 0)}
\nonumber\\
&=&\frac{\omega_2}
{\omega_2+\Sigma^{H}_{N+1}-\Sigma^{H}_{N}}> 1.
\label{static_uniform_dielectric}
\end{eqnarray}
Note that $\epsilon_{N,N+1}(0,\vec{0}\,)$ 
does not go to unity because of the finite 
direct (Hartree) self-energy term, showing a 3D property.
In a ZWW, however, the Hartree self-energy is a constant independent
of the level index, and therefore is cancelled with each other in 
Eq. (\ref{static_uniform_dielectric}).
When taking the large momentum limit ($|\vec{q}_\perp|\rightarrow\infty$),
$\epsilon_{N,N+1}(0,\vec{q}_\perp)\rightarrow 1$ 
for $\omega_\rho(\vec{q}_\perp)-\omega_{\sigma_z}(\vec{q}_\perp)\to 0$.

As in the ZWW, 
a scalar dielectric function similar to Eq. (\ref{dielectric_zerowidth})
can be obtained for a WWW
system subject to a strong in-plane magnetic field.
The similarity between these two systems is because 
the strong in-plane magnetic field effectively enhances the
electron confinement energy of the well (note that
$\omega_b=\sqrt{\omega_0^2+\omega_\|^2}$ and see Sec. \ref{noninteracting}). 
We start from the following general approximation
(we use number labels (e.g. 1,2 $\cdots$) and Greek labels
(e.g. $\alpha$, $\beta$ $\cdots$) to replace the level indices,
$n_{1,2\cdots}$ and $m_{\alpha,\beta\cdots}$ for simplicity):
\begin{eqnarray}
&&U_{14,\alpha\beta}(\vec{q}_\perp)U_{\mu\nu,23}(\vec{q}_\perp)
\nonumber\\
&=&\frac{1}{L_z^2}\sum_{q_z'}A_{14}(-\vec{q}_\perp,-q_z')V(\vec{q}_\perp,q_z')
A_{\alpha\beta}(\vec{q}_\perp,q_z')
\sum_{q_z"}A_{\mu\nu}(-\vec{q}_\perp,-q_z")V(\vec{q}_\perp,q_z")
A_{23}(\vec{q}_\perp,q_z")
\nonumber\\
&=&\frac{1}{L_z^2}\sum_{q_z'}A_{14}(-\vec{q}_\perp,-q_z')V(\vec{q}_\perp,q_z')
A_{23}(\vec{q}_\perp,q_z')
\sum_{q_z"}A_{\mu\nu}(-\vec{q}_\perp,-q_z")V(\vec{q}_\perp,q_z")
A_{\alpha\beta}(\vec{q}_\perp,q_z")
\times\underbrace{\frac{A_{23}(\vec{q}_\perp,q_z")}{A_{23}(\vec{q}_\perp,q_z')}
\frac{A_{\alpha\beta}(\vec{q}_\perp,q_z')}{A_{\alpha\beta}(\vec{q}_\perp,q_z")}
}_{\displaystyle\equiv C^{23}_{\alpha\beta}(\vec{q}_\perp,q_z',q_z")}
\nonumber\\
&\sim& U_{14,23}(\vec{q}_\perp)U_{\mu\nu,\alpha\beta}(\vec{q}_\perp),
\label{UU_approx}
\end{eqnarray}
where $C^{23}_{\alpha\beta}(\vec{q}_\perp,q_z',q_z")$ has been approximated
by its zeroth order value
\begin{eqnarray}
C^{23}_{\alpha\beta}(\vec{q}_\perp,q_z',q_z")\sim
1+O(\omega_\perp/\omega_b),
\label{C_function}
\end{eqnarray}
according to the explicit expression of $A_{\alpha\beta}(\vec{q})$ shown
in Appendix \ref{A_function}.
Therefore the TDHFA screening similar to Eq. (\ref{dielectric_zerowidth})
for a WWW system
could be obtained approximately as:
\begin{eqnarray}
\varepsilon(\omega,\vec{q}_\perp)&\sim&
\left(1-\sum_{\alpha\beta}\sum_{\mu \nu}\left[
Y_{\alpha \beta,\mu \nu}(\omega,\vec{q}_\perp)\right]^{-1}
U_{\mu\nu,\alpha\beta}(\vec{q}_\perp) \right)^{-1}.
\label{dielectric_finitewidth}
\end{eqnarray}
Similarly the high density approximation with RPA diagrams only can
also be obtained by using the same approximation:
\begin{eqnarray}
&&\varepsilon_{RPA}(\omega,\vec{q}_\perp)\sim 1+
\frac{1}{2\pi l_0^2}\sum_{\sigma}
\sum_{m=0}^{N_{\sigma}}\sum_{n=N_{\sigma}+1}^{\infty}
\frac{2(E_{n,\sigma}-E_{m,\sigma})\,\omega_2}
{(E_{n,\sigma}-E_{m,\sigma})^2\omega^2_2+\omega^2}
\cdot\frac{1}{L_z}\sum_{p_z} V(q_x,q_y,p_z)
\frac{m!}{n!}
\nonumber\\
&&\times
\exp\left[-\frac{\cos^2\theta(q_yl_0)^2+
(\cos\theta q_xl_0-\sin\theta p_zl_0)^2\lambda_1^2}{2\lambda_1}\right]
\exp\left[-\frac{\sin^2\theta(q_yl_0)^2+
(\sin\theta q_xl_0+\cos\theta p_zl_0)^2\lambda_2^2}{2\lambda_2}\right]
\nonumber\\
&&\times
\left(\frac{\sin^2\theta(q_yl_0)^2+
(\sin\theta q_xl_0+\cos\theta p_zl_0)^2\lambda^2_2}{2\lambda_2}\right)
^{n-m}
\left[L_{m}^{n-m}
\left(\frac{\sin^2\theta(q_yl_0)^2+
(\sin\theta q_xl_0+\cos\theta p_zl_0)^2\lambda_2^2}{2\lambda_2}\right)\right]^2
.
\label{full_expression}
\end{eqnarray}
Comparing results of Eq. (\ref{pi_aleiner}) for a ZWW and Eq. 
(\ref{full_expression}) for a wide (parabolic) well, we find that the
finite width effect enhances the anisotropy of the dielectric function
through the coupling of $x$ and $z$ components of wavevectors.
Note that Eqs. (\ref{static_finitewidth}), (\ref{dielectric_finitewidth}),
and (\ref{full_expression}) show no screening in the
$z$ direction because we have integrated out the $z$ component
in the interaction matrix element by
the single particle wavefunctions in Eq. (\ref{bare_matrix})
and have assumed the level index dependence of the dielectric function
to be unimportant (see Eq. \ref{C_function})). We believe that this is a good
approximation for strong in-plane magnetic fields
(see Fig. \ref{energy_levels_figure}),
so that there is no appreciable static or dynamical
polarization in the $z$ direction to screen the Coulomb interaction.
This approximation certainly fails when one
wants to study excitations between levels of two different subbands in a
weak in-plane field region.
\subsection{Numerical Results}

\begin{figure}
 \vbox to 5cm {\vss\hbox to 5.5cm
 {\hss\
   {\includegraphics{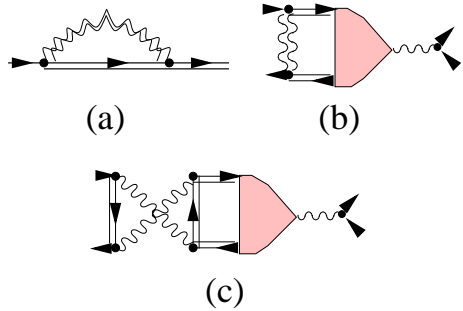}
   }
  \hss}
 }
\caption{
(a) The screened exchanged energy and (b) the screened ladder diagrams
used in the screened TDHFA developed in Sec. \ref{screening}
(see Fig. \ref{feynm_figure}). The
interaction lines of the direct
(Hartree) energy and the RPA diagrams are not screened to avoid
double counting.
(c) A diagram not included in the screened TDHFA but of the
same order as a screened ladder diagram shown in (b).
All notations are the same as those in Fig. \ref{feynm_figure}.
}
\label{other_diagram}
\end{figure}

In this section, we show some numerical results of the
collective mode energies including the screening effect.
For convenience, we choose the dielectric function shown 
in Eq. (\ref{full_expression})
and consider static screening ($\omega=0$) only. Therefore the algebraic
matrix equations of Eqs. (\ref{U_screened})-(\ref{Y_matrix}) are
all of the same form except the Coulomb interaction is replaced by the
screened one, $V(\vec{q}\,)/\epsilon_{RPA}(\vec{q}_\perp,0)$. 
However, the interaction of the RPA energy
in Eq. (\ref{U_RPA}) and the Hartree self-energy are not 
screened in order to avoid double counting of bubble diagrams
(see Figs. \ref{other_diagram}(a) and (b)). We note that such screened TDHFA
is {\it not} a strictly current-conserving approximation, because
some other diagrams (for example, see Fig. \ref{other_diagram}(c)) 
are not included, 
which may contribute to the same higher order effects
as the screening bubbles.
Therefore we can only estimate the screening
effect to the magnetoplasmon energy qualitatively rather than quantitatively
in our present study \cite{footnote}.

\begin{figure}
 \vbox to 6cm {\vss\hbox to 5.5cm
 {\hss\
    {\includegraphics{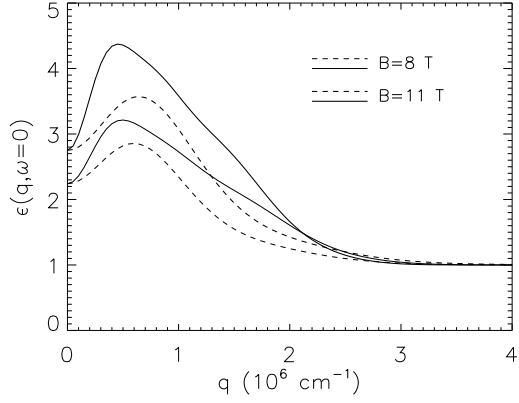}
  }
  \hss}
  }
\caption{
Static dielectric function in momentum space for $\nu=6$ and $\omega_0=7$ meV.
Solid and dashed lines represent 
$\epsilon(q_x,q_y=0)$ and $\epsilon(q_x=0,q_y)$ respectively. Thick and thin
lines are for $B_\|=11$ and $8$ T.
}
\label{dielectric}
\end{figure}

In the presence of screening, 
the first order phase transition point, $B_\|^\ast$, moves higher
values (see the fourth column of Table \ref{critical_B_table}) because
the exchange interaction strength is reduced.
This allows us to investigate the 
magnetoplasmon mode dispersion at higher values of in-plane magnetic field
without changing the ground state configuration (i.e. avoiding the trivial
first-order transition).
In Fig. \ref{dielectric} we show the static dielectric function, 
$\epsilon(\omega=0,\vec{q}_\perp)$ obtained by
Eq. (\ref{full_expression}) in RPA for two different
values of in-plane magnetic field at $\nu=6$. 
For a stronger in-plane field,
the screening effect is also stronger and more anisotropic.
The anisotropic dielectric function shows
that interaction along $x$ direction (parallel to the in-plane field)
is screened more than the interaction along $y$ direction (perpendicular 
to the in-plane field).

\begin{figure}
 \vbox to 10cm {\vss\hbox to 5.5cm
 {\hss\
    {\includegraphics{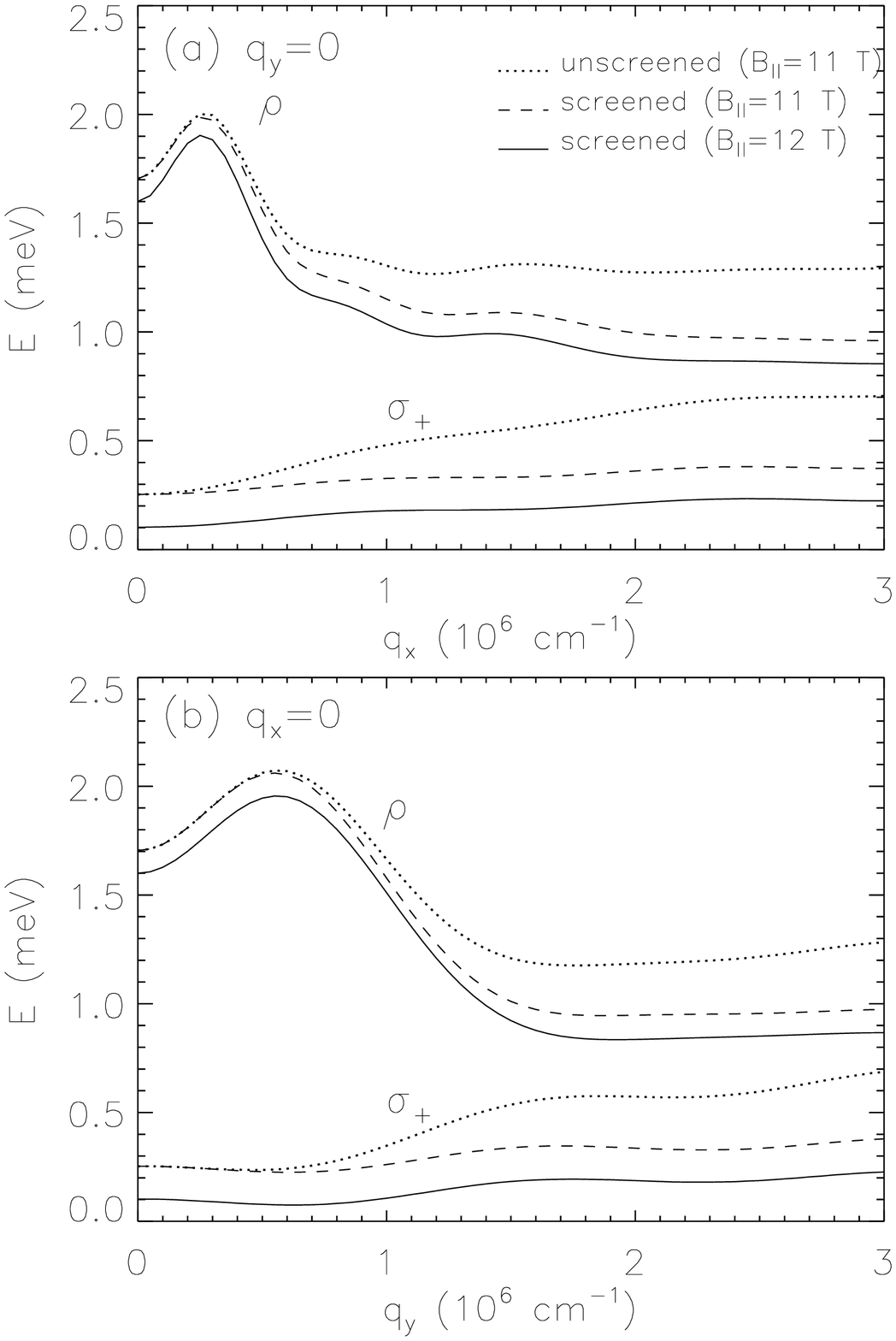}
  }
  \hss}
  }
\caption{
Dispersions of magnetoplasmon excitations for $\nu=6$ in both charge ($\rho$)
and spin ($\sigma_+$) modes 
including RPA screening (Eq. (\ref{full_expression})) of the 
Coulomb interaction for $B_\|=11$ and 12 Tesla (dashed and solid 
lines respectively).
Results of unscreened dispersion are also shown 
(dotted lines, the same as Fig. \ref{unscrn_2l_cs_n6_w7})) for comparison.
}
\label{pl_screened}
\end{figure}

In Fig. \ref{pl_screened} we show 
calculational results of the charge 
($\omega_\rho$) and spin ($\omega_{\sigma_+}$) 
mode magnetoplasmon dispersions including RPA screening effects (dashed lines)
with other parameters the same as Fig. \ref{unscrn_2l_cs_n6_w7}.
For comparison, the unscreened results 
(dotted lines) and the screened results with higher 
in-plane magnetic field (solid lines) are shown together in the same figure.
Comparing unscreened and screened results at $B_\|=11$ Tesla
(dotted and dashed lines respectively),
one can find that the screening effect does lower 
magnetoplasmon energies for both charge and spin modes in the
large $|\vec{q}_\perp|$ region due to the shrinking
of Fock self-energy. But this effect is relatively
weaker in the long wavelength limit (small $|\vec{q}_\perp|$ limit) due to
the cancelation between the Fock self-energy and the electron-hole binding
energy (the generalized Kohn's theorem). In the intermediate $|\vec{q}_\perp|$
region, the roton-minimum becomes less prominent than the unscreened result,
and the dispersion becomes flat. 
Therefore, fixing all the other system parameters, the screening
effect is not very important in determining the roton-minimum energy.
On the other hand, as mentioned above, the screening effect 
reduces the electron self-energy and increases the critical value of $B_\|$
for the unpolarized-to-polarized first order 
phase transition. Therefore one can, in the presence of screening,
calculate the screened collective mode at higher in-plane magnetic field
based on the same unpolarized ground state since the first order transition
is now pushed to higher fields.
In Fig. \ref{pl_screened}
we show the result of magnetoplasmon dispersion calculated at $B_\|=12$ Tesla
(solid lines). The roton minimum of the spin collective mode ($\sigma_+$)
becomes lower than 0.1 meV, showing an almost mode softening at finite
wavevector along the direction perpendicular to the in-plane magnetic field.
Our results therefore indicate that inclusion of screening,
effect, as well as lowering the confinement potential and increasing the
electron density, could help to stabilize a new anisotropic ground 
state with broken translational
and spin symmetries associated with the softening of the spin collective mode.
Such a symmetry broken phase may very well be the cause for transport
anisotropy observed in Ref. \cite{pan01}.

\section{discussion}
\label{discussion}

In this section, we briefly discuss the possible phases of this
new ground state based on our collective mode calculation
results shown above. More detailed 
theoretical results on these exotic quantum phases will be 
given elsewhere \cite{unpublished}.
Similar to a DQW system \cite{eugene01}, where 
the layer index is treated as an isospin degree of freedom, 
the level index of the closest two levels around the Fermi level
can be used to construct an isospin (here it is also one-to-one 
related to a real spin)
space, and create a coherent wavefunction
for the possible new ground state
in a single Slater determinant, 
\begin{eqnarray}
|\Psi_1 \rangle =
\prod_{k}
(~ e^{ikQ_xl_0^2} \,
\cos \frac{w_k}{2} c_{N,k-Q_y/2,\downarrow}^{\dagger}
+ e^{-ikQ_xl_0^2} \,
\sin \frac{w_k}{2} c_{N+1,k+Q_y/2,\uparrow}^{\dagger} ~)~
|0 \rangle,
\label{new_groundstate_wavefunction}
\end{eqnarray}
where $c^\dagger_{n,k,s}$ creates an electron in state
$\phi_{n,k,s}(\vec{r}\,)$ with spin $s$, and $|0 \rangle$ 
denotes the ground state with $N+1$ filled Landau levels of spin up and
$N$ levels of spin down.
We consider six different phases constructed from 
Eq. (\ref{new_groundstate_wavefunction}), corresponding to 
different variational parameters, $w_k$ and $\vec{Q}$. 
When  $w_k$ is constant, the wavefunction of Eq. 
(\ref{new_groundstate_wavefunction}) can describe three non-stripe
phases: (i) a fully (un)polarized uniform quantum Hall 
phases for $w_k=(0)\pi$, (ii)
a simple inter-level coherent phase for $\vec{Q}=0$ and $w_k\neq 0,\pi$, and
(iii) a spiral phase for finite $\vec{Q}$ and $w_k\neq 0,\pi$.
When $w_k$ changes periodically with $k$, three
different kinds of stripe phases arise: (i)
simple stripe phase for $\vec{Q}=0$, which has no spiral structure, 
(ii) skyrmion stripe phase for finite $\vec{Q}$, but $\vec{Q}\perp\hat{n}$,
where $\hat{n}$ is the normal vector of the stripe formation.
Such a skyrmion stripe phase has both charge
and spin modulation in different directions, and therefore has finite 
topological charge density oscillation in real space 
\cite{eugene01,unpublished,sondhi93};
(iii) spiral stripe phase for finite $\vec{Q}$ with $\vec{Q}\,\|\,\hat{n}$.
This spiral stripe phase has charge and spin modulation in the 
same direction but no topological charge oscillation in real space. 
We should point out that the wavefunction 
of Eq. (\ref{new_groundstate_wavefunction})
is based on a special choice of Landau gauge, 
$\vec{A}=(0,B_\bot x-B_\| z,0)$, and therefore
gives the stripe direction along $y$, i.e. perpendicular to the 
in-plane magnetic field. Choosing another kind of Landau
gauge, where electron momentum is conserved along $x$, 
$\vec{A}=(-B_\perp y,-B_\| z,0)$, we can construct a stripe along $x$
direction and the noninteracting Hamiltonian can be solved exactly
by a canonical transformation \cite{unpublished}. 
Then one can write the trial wavefunctions of these different phases
and obtain their energies in Hartree-Fock approximation. The one of the lowest
energy states should be the ground state near the degeneracy point,
$B_\|=B_\|^\ast$. Details will be presented elsewhere \cite{unpublished}.

On the other hand,
the magnetoplasmon excitation spectra we obtain in previous sections
also gives us 
important information about the new ground state near the 
degeneracy point. 
First, the asymmetry of spin 
density mode in $x$ and $y$ direction and the near mode softening
in $y$ direction (shown in Fig. \ref{unscrn_2l_cs_n6_w7})
strongly indicate that the new symmetry-broken ground state, 
if it exists, should
have a spin spiral structure at finite wavevector in $y$ direction.
This may be a spiral spin density wave, when only one of the ordering
wavevectors $\pm (0, q_y^\ast)$ is present, or a collinear
spin density wave, when there is ordering at both wavevectors
with equal amplitudes. The former can be visualized as a spin density wave
where electron spin has a spiral structure around 
the total magnetic field direction
in order to optimize the exchange energy.
Therefore collinear spin density wave, spiral, skyrmion stripe, and
spiral stripe phases are the possible candidates for the symmetry-broken
phase. 
As for the existence of any possible charge 
density wave instability, we could not obtain 
much information from our collective mode calculation in
TDHFA. But it is apparently true that interaction effects are more
important for stronger
in-plane magnetic fields, where the noninteracting energy 
separation, $\omega_2$, becomes very small.

Considering the experimental results \cite{pan01}, where
the resistance along the in-plane magnetic field
becomes finite when the in-plane magnetic field exceeds
a critical value, we find that the stripe
formation, if it exists, should be along the direction perpendicular to
the in-plane magnetic field (i.e. its normal wavevector is 
in $x$ direction) to produce such a transport
anisotropy. Using the above result that the spin density modulation has a 
wavevector, $q_y^\ast$, in $y$ direction, we find that 
only the skyrmion stripe phase is consistent with all of these constraints
and should be the best candidate for the new ground state. 
Although our Hartree-Fock calculation
shows that the spiral phase has slightly lower energy than
the skyrmion stripe phase \cite{eugene01}, 
we believe that this may be due to the nonparabolicity of a realistic
WWW or the correlation effects not included in our HF approximation.
We therefore speculate that the anisotropic ground state observed
in Ref. \cite{pan01} is our proposed novel spin skyrmion stripe phase.
This may also be true for the transport anisotropy earlier observed
\cite{zeitler} in Si based 2D systems, but the additional complications
of valley degeneracy in Si makes the application of our theory mode different.

\section{Summary}
\label{summary}

We study the magnetoplasmon excitations of a parabolic
quantum well system in a tilted magnetic field. Starting from
the many-body theory in coordinate space, we integrate out the continuous
variable and obtain an algebraic matrix representation of the dielectric
function and hence the magnetoplasmon mode dispersion in TDHFA.
Focusing on even filling factors, a 
roton-minimum near zero energy in the spin channel is observed
at finite wavevector along the direction 
perpendicular to the in-plane magnetic field. 
By changing the confinement potential, we have a continuous 
transition from a 3D plasmon excitation to the pure 2D results
in our calculation.
Including the screening effect, which is another important 
part of our work, we
find that the roton-minimum energy could be even more 
suppressed. Although it does not
reach zero energy before possibly undergoing 
a first order phase transition from 
an unpolarized ground state to a polarized one, 
its small excitation energy
at finite wavevector suggests a possible spin-density-instability  
to an exotic symmetry-broken ground state in realistic systems.
We discuss various phases that may result and propose that
the recent transport anisotropy measurement in experiments \cite{pan01}
can be explained by a skyrmion stripe phase, where spin
and charge density modulations are in different directions. 
The theoretical technique used in this paper could also be used
to study other quantum Hall systems in quasi-2D quantum well
nanostructures. In particular, our screening theory is more 
complete than the existing theory, and should have wide 
applicability. Finally we point out that our predicted 
collective mode dispersion may be directly verified via 
the inelastic light scattering spectroscopy.

\section{Acknowledgments}
\label{acknowledgement}

This work is supported by the NSF-DMR (E.D. and B.I.H) , US-ONR
(D.W.W. and S.D.S.), and the Harvard
Society of Fellows (E.D.). We acknowledge useful discussions with
C. Kallin, S. Kivelson, A. Lopatnikova, A. MacDonald, I. Martin,
C. Nayak, L. Radzihovsky, and S. Simon.
\appendix

\section{Self-Consistent Hartree-Fock Equations}
\label{HF_equation_append}

In this section we derive the self-consistent 
Hartree-Fock equations for the
single particle wavefunctions. Starting from Eq. (\ref{HF_equation_general}),
we can use a Fourier transform of $V(\vec{r}\,)$ to obtain
\begin{eqnarray}
{E}_{\vec{n},\sigma}{\phi}_{\vec{n},k,\sigma}(\vec{r}\,)
&=&\left[{\cal H}_0+\frac{1}{\Omega}\sum_{\vec{q}}V(\vec{q})
\sum_{\vec{m},p,s}\nu_{m,p,s}\int d\vec{r}\,'e^{-i\vec{q}\cdot\vec{r}\,'}
{\phi}_{\vec{m},p,s}{}^\dagger(\vec{r}\,')
{\phi}_{\vec{m},p,s}(\vec{r}\,')e^{i\vec{q}\cdot\vec{r}}
\right]{\phi}_{\vec{n},k,\sigma}(\vec{r}\,)
\nonumber\\
&&-\frac{1}{\Omega}\sum_{\vec{q}}V(\vec{q})
\int d\vec{r}\,'e^{-i\vec{q}\cdot\vec{r}\,'}
{\phi}_{\vec{n},k,\sigma}(\vec{r}\,')
\sum_{\vec{m},p}\nu_{\vec{m},p,\sigma}
{\phi}_{\vec{m},p,\sigma}{}^\dagger(\vec{r}\,')
e^{i\vec{q}\cdot\vec{r}}{{\phi}_{\vec{m},p,\sigma}}(\vec{r}\,)
\nonumber\\
&=&\left[{\cal H}_0+\frac{1}{\Omega}\sum_{\vec{q}}V(\vec{q})
\sum_{\vec{m},p,s}\nu_{m,p,s}\delta_{q_y,0}e^{i(p+q_y/2)q_xl_0^2}
{A}_{\vec{m}\vec{m}}^{s,s}(\vec{q}\,)e^{i\vec{q}\cdot\vec{r}}\right]
{\phi}_{\vec{n},k,\sigma}(\vec{r}\,)
\nonumber\\
&&-\frac{1}{\Omega}\sum_{\vec{q}}V(\vec{q})
\sum_{\vec{m},p}\nu_{\vec{m},p,\sigma}\delta_{k-p,q_y}e^{i(p+q_y/2)q_xl_0^2}
{A}_{\vec{m},\vec{n}}^{\sigma,\sigma}(\vec{q}\,)
e^{i\vec{q}\cdot\vec{r}}\phi_{\vec{m},p,\sigma}(\vec{r}\,),
\end{eqnarray}
where the form function, ${A}_{\vec{m},\vec{n}}^{\sigma,\sigma}(\vec{q}\,)$,
has been defined in Eq. (\ref{A_def}). 
Assuming a uniform ground state, we can separate 
${\phi}_{\vec{n},k,\sigma}(\vec{r})$ into a product of a plane wave,
$e^{iky}/\sqrt{L_y}$, and the function,
${\Phi}_{\vec{n},\sigma}(x+kl_0^2,z)$, which
satisfies the following eigenvalue equation:
\begin{eqnarray}
{E}_{\vec{n},\sigma}{\Phi}_{\vec{n},\sigma}(x,z)
&=&\left[{\cal H}_0+
\frac{1}{2\pi l_0^2 L_z}\sum_{q_z}V(q_z)\sum_{\vec{m},\sigma'}
\nu_{\vec{m},\sigma'}
{A}_{\vec{m}\vec{m}}^{\sigma'\sigma'}(q_z)\,e^{iq_z z}\right]
{\Phi}_{\vec{n},\sigma}(x,z)
\nonumber\\
&&-\frac{1}{\Omega}\sum_{\vec{q}}V(\vec{q}\,)
\sum_{\vec{m}}\nu_{\vec{m},\sigma}\,e^{-iq_xq_y/2}
{A}_{\vec{m}\vec{n}}^{\sigma\sigma}(\vec{q}\,)\,e^{iq_xx+iq_zz}
{\Phi}_{\vec{m},\sigma}(x-q_yl_0^2,z),
\label{selfconsistent_HF}
\end{eqnarray}
where $\nu_{\vec{m},\sigma}$ 
is the filling factor of Landau level $\vec{m}$ and
spin $\sigma$, satisfying
\begin{eqnarray}
\nu=\sum_{\vec{m},\sigma}\nu_{\vec{m},\sigma},
\end{eqnarray}
to conserve the total electron density.

Now we expand ${\Phi}_{\vec{n},\sigma}(x,z)$ in terms of noninteracting
wavefunctions with the same spin (note that $V(\vec{r}\,)$ 
allows no spin-flip,
so that $\sigma$ is conserved and no spin hybridization occurs):
\begin{eqnarray}
{\Phi}_{\vec{n},\sigma}(x,z)=\langle x,z|{\vec{n},\sigma}\rangle
=\sum_{\vec{m}}\langle x,z|\vec{m},\sigma\rangle_0\,
{}_0\langle\vec{m},\sigma |{\vec{n},\sigma}\rangle,
\end{eqnarray}
where $|\cdots\rangle_0$ represents a noninteracting eigenstate. We have
\begin{eqnarray}
{A}^{\sigma\sigma}_{\vec{m}\vec{n}}(\vec{q}\,)
&=&\int dx\int dz\,e^{-iq_xx-iq_zz}
{\Phi}^\dagger_{\vec{m},\sigma}(x-q_yl_0^2/2,z)
{\Phi}_{\vec{n},\sigma}
(x+q_yl_0^2/2,z) \nonumber\\
&=&\sum_{\vec{l}_1}\sum_{\vec{l}_2}
\langle{\vec{m},\sigma}|\vec{l}_1,\sigma\rangle_0\,
{}_0\langle\vec{l}_2,\sigma |{\vec{n},\sigma}\rangle
A^{(0),\sigma\sigma}_{\vec{l}_1,\vec{l}_2}(\vec{q}\,).
\label{A_tilde}
\end{eqnarray}
Using Eq. (\ref{A_tilde}) and multiplying by 
the noninteracting wavefunction from
the left of Eq. (\ref{selfconsistent_HF}), we have the self-consistent
Hartree-Fock equation in a matrix representation:
\begin{eqnarray}
&&{E}_{\vec{n},\sigma}
{}_0\langle\vec{n}\,',\sigma |{\vec{n},\sigma}\rangle
\nonumber\\
&=&\sum_{\vec{m}\,'}\left[E^0_{\vec{m}\,',\sigma}\delta_{\vec{n}\,',\vec{m}\,'}
+\frac{1}{2\pi l_0^2 L_z}\sum_{q_z}V(q_z)\sum_{\vec{m},\sigma'}
\nu_{\vec{m},\sigma'}\sum_{\vec{l}_1}\sum_{\vec{l}_2}
\langle{\vec{m},\sigma'}|\vec{l}_1,\sigma'\rangle_0\,
{}_0\langle\vec{l}_2,\sigma' |{\vec{m},\sigma'}\rangle
A^{(0),\sigma'\sigma'}_{\vec{l}_1,\vec{l}_2}(q_z)
A^{(0),\sigma\sigma}_{\vec{n}\,',\vec{m}\,'}(-q_z)\right]
{}_0\langle\vec{m}\,',\sigma |{\vec{n},\sigma}\rangle
\nonumber\\
&&-\sum_{\vec{m}\,'}\frac{1}{\Omega}\sum_{\vec{q}}V(\vec{q}\,)
\sum_{\vec{m}}\nu_{\vec{m},\sigma}
\sum_{\vec{l}_1}\sum_{\vec{l}_2}
\langle{\vec{m},\sigma}|\vec{l}_1,\sigma\rangle_0\,
{}_0\langle\vec{m}\,',\sigma |{\vec{n},\sigma}\rangle
A^{(0),\sigma\sigma}_{\vec{l}_1,\vec{m}\,'}(\vec{q}\,)
A^{(0),\sigma\sigma}_{\vec{n}\,',\vec{l}_2}(-\vec{q}\,)
\,{}_0\langle\vec{l}_2,\sigma |{\vec{m},\sigma}\rangle
\nonumber\\
&=&\sum_{\vec{m}\,'}\left[E^0_{\vec{m}\,',\sigma}\delta_{\vec{n}\,',\vec{m}\,'}
+\sum_{\vec{m},\sigma'}\nu_{\vec{m},\sigma'}
\sum_{\vec{l}_1}\sum_{\vec{l}_2}
\langle{\vec{m},\sigma'}|\vec{l}_1,\sigma'\rangle_0\,
{}_0\langle\vec{l}_2,\sigma' |{\vec{m},\sigma'}\rangle
\left\{U^{(1),\sigma\sigma'}_
{\vec{n}\,',\vec{m}\,';\vec{l}_1,\vec{l}_2}(\vec{0}_\perp)
-U^{(1),bind,\sigma\sigma}_
{\vec{n}\,',\vec{l}_2;\vec{l}_1,\vec{m}\,'}(\vec{0}_\perp)
\delta_{\sigma,\sigma'}\right\}\right]
{}_0\langle\vec{m}\,',\sigma |{\vec{n},\sigma}\rangle,
\nonumber\\
\label{selfconsistent_HF2}
\end{eqnarray}
where we have used Eq. (\ref{U_explicit}) and Eq. (\ref{U'_explicit}) to express
the direct and the exchange potential. Eq. (\ref{selfconsistent_HF2}) is the 
matrix representation of the Hartree-Fock Hamiltonian in our system, which
should be solved self-consistently to get the energy eigenstate via
vector elements, ${}_0\langle\vec{m},\sigma |{\vec{n},\sigma}\rangle$. 

Another expression for the eigenenergies can be obtained directly from
Eq. (\ref{selfconsistent_HF}), by integrating another eigenket
${\Phi}_{\vec{n},\sigma}(x,z)$ from the left. We obtain
\begin{eqnarray}
{E}_{\vec{n},\sigma}
&=&E^0_{\vec{m}\,',\sigma}+
\frac{1}{2\pi l_0^2 L_z}\sum_{q_z}V(q_z)\sum_{\vec{m},\sigma'}
\nu_{\vec{m},\sigma'}
{A}_{\vec{m}\vec{m}}^{\sigma'\sigma'}(q_z)
{A}_{\vec{n}\vec{n}}^{\sigma\sigma}(-q_z)
-\frac{1}{\Omega}\sum_{\vec{q}}V(\vec{q}\,)
\sum_{\vec{m}}\nu_{\vec{m},\sigma}
{A}_{\vec{m}\vec{n}}^{\sigma\sigma}(\vec{q}\,)
{A}_{\vec{n}\vec{m}}^{\sigma\sigma}(-\vec{q}\,),
\nonumber\\
&=&E^0_{\vec{m}\,',\sigma}+\sum_{\vec{m},\sigma'}\nu_{\vec{m},\sigma'}
\left\{{U}^{\sigma\sigma'}
_{\vec{n},\vec{n};\vec{m},\vec{m}}(\vec{0}_\perp)
-{U}^{bind,\sigma\sigma}_{\vec{n},\vec{m};\vec{m},\vec{n}}(\vec{0}_\perp)
\delta_{\sigma,\sigma'}\right\},
\label{SCHF_energy}
\end{eqnarray}
where ${U}^{\sigma\sigma'}
_{\vec{n},\vec{n};\vec{m},\vec{m}}(\vec{0}_\perp)$ and 
${U}^{bind,\sigma\sigma}_{\vec{n},\vec{m};\vec{m},\vec{n}}
(\vec{0}_\perp)$ are those defined in 
Eq. (\ref{U_bind}) and Eq. (\ref{U_RPA}).

Using this self-consistent Hartree-Fock equation, Eq. 
(\ref{selfconsistent_HF2}), it is also easy
to include any nonparabolic effects of the realistic confinement
potential, $U(z)$. 
Assuming the deviation of the realistic $U(z)$ from 
a parabolic one, $U_p(z)$,
to be small, i.e. $|\Delta U(z)=U(z)-U_p(z)|\ll \omega_0$,
we can calculate its matrix element,
\begin{eqnarray}
\langle \vec{n}\,',\sigma|\Delta U(z)|
\vec{m}\,',\sigma\rangle
&=&\int dx\int dz\, \Phi_{\vec{n}\,'}(x,z)\Delta U(z)\Phi_{\vec{m}\,'}(x,z)
\nonumber\\
&=&\int dx\int dz\, \Phi_{\vec{n}\,'}(x,z)\left[\frac{1}{L_z}\sum_{q_z}
\Delta U(q_z)\,e^{iq_zz}\right]\Phi_{\vec{m}\,'}(x,z)
\nonumber\\
&=&\frac{1}{L_z}\sum_{q_z}\Delta U(q_z)A_{\vec{n}\,'\vec{m}\,'}(\vec{0}_\perp,
-q_z)
=\frac{1}{2\pi}\int dq_z\Delta U(q_z)A_{\vec{n}\,'\vec{m}\,'}(\vec{0}_\perp,
-q_z).
\end{eqnarray}
and incorporate it in Eq. (\ref{selfconsistent_HF2}) 
to calculate the self-consistent
Hartree-Fock eigenenergies and eigenfunctions.
In all our numerical work presented in this paper, 
however, we have taken $U(z)$ to be parabolic
throughout.
\section{Magnetoplasmon Excitation Energy through the Magnetic Exciton
Wavefunction} 
\label{realspace_expression}

In this section we show that
the magnetoplasmon excitation energies both
in a thin 2D (ZWW) well in only a perpendicular magnetic
field (situation discussed in \cite{kallin}) and in
a wide parabolic well with a tilted magnetic field
(situation discussed in this paper)
can be written in a simple and instructive
form by using exciton wavefunctions proposed
in \cite{kallin} and its appropriate WWW generalization constructed in our
Eq. (\ref{wavefunction-WW}), respectively. 
For the first case,
we take the static exciton wavefunction 
suggested by Kallin and Halperin in Eq. (2.9) of Ref. \cite{kallin}
and set the center of mass coordinate and the total momentum of excitons
to be zero:
\begin{equation}
\Psi^{2D}_{n_\beta,n_\alpha}(\Delta x,\Delta y)\equiv
\int d\eta\, e^{-i\eta\Delta y/l_0^2}
{\psi}^{(0)}_{n_\beta}(\eta+\Delta x/2)
{\psi}^{(0)}_{n_\alpha}(\eta-\Delta x/2),
\label{wavefunctions-KH}
\end{equation}
where $\Delta x$ and $\Delta y$ are the relative coordinates between
the hole in a filled level (denoted by $n_\alpha=n$) and 
the electron in an empty level (denoted by $n_\beta=n_\alpha+m$);
${\psi}^{(0)}_{n}(x)$ is the wavefunction of one-dimensional single
harmonic oscillator as shown in Eq. (\ref{wf_i}) with $l_i$ replaced by $l_0$.
In the lowest order of $(e^2/\epsilon l_0)/\omega_\bot$, there are
four distinct contribution to the  
magnetoplasmon excitation energies:
noninteracting energy separation,
exciton binding energy, RPA energy, and exchange self-energy \cite{kallin},
\begin{equation}
\omega^{2D}_{mn,\rho}({q})=
m\omega_\perp-\omega_z(\sigma_\beta-\sigma_\alpha)+
\Delta E^{mn}_{bind}({q})+
\Delta E^{mn}_{RPA}({q})+\Delta E^{mn}_{exch},
\label{E-KH}
\end{equation}
where the last three terms can be re-expressed in terms of
$\Psi^{2D}_{n_\beta,n_\alpha}$ as follows
\begin{eqnarray}
\Delta E^{mn}_{bind}({q})&=&-\frac{1}{2\pi l_0^2}
\int d\Delta\vec{r}_\perp
V^{2D}(\Delta\vec{r}-l_0^2\vec{q}_\perp\times\hat{z})\left|\Psi^{2D}_{n+m,n}
(\Delta \vec{r}_\perp)\right|^2
\label{KH-bind}\\
\Delta E^{mn}_{RPA}({q})&=&\frac{2V^{2D}({q})}{2\pi l_0^2}
\left|\Psi^{2D}_{n+m,n}(-q_yl_0^2,q_xl_0^2)\right|^2
\label{KH-RPA}\\
\Delta E^{mn}_{exch}&=&\Sigma_{n+m}^{F}-\Sigma_n^{F}=
\frac{-1}{2\pi l_0^2}\int d\Delta\vec{r}_\perp
V^{2D}(\Delta\vec{r}_\perp)\left[\Psi^{2D}_{n+m,n+m}(\Delta \vec{r}_\perp)
\sum_{l\leq N_{\sigma_\beta}}{\Psi^{2D}_{l,l}}^\ast(\Delta \vec{r}_\perp)
\right.
\nonumber\\
&&\hspace{2.5cm}\left.-\Psi^{2D}_{n,n}(\Delta \vec{r}_\perp)
\sum_{l\leq N_{\sigma_\alpha}}
{\Psi^{2D}_{l,l}}^\ast(\Delta \vec{r}_\perp) \right],
\label{KH-exchange}
\end{eqnarray}
where
$N_{\sigma_{\alpha(\beta)}}$ is the level index of the highest
occupied Landau level with spin $\sigma_{\alpha(\beta)}$.
Interpretation of the  formulas in (\ref{KH-bind}) and (\ref{KH-RPA}) is
straightforward. The binding energy integrates over relative
positions of electron and hole in the exciton, whereas
the RPA
term involves electron and hole annihilating each other
and is proportional to the probability of finding two particles
at the same position.
$\Delta E^{mn}_{exch}$ in Eq. (\ref{KH-exchange}) is the difference of
exchange self-energies between the two relevant levels, and
indicates the relative many-body level shift. The exchange self-energy
of level $n$, $\Sigma^{F}_n$, expressed in Eq. (\ref{KH-exchange}) can be
understood as the integral over relative positions of electrons
between level $n$ and the lower levels, $l$, of the same spin. 
Note that $\omega^{2D}_{mn}(\vec{q}_\perp=0)=m\omega_\perp$ for
$m=1$ in the charge mode channel, 
satisfying Kohn's theorem \cite{kohn} for this
ZWW system.
The equivalence between above expressions and the results in \cite{kallin}
can be easily seen by direct substitution. 

For a parabolic well,
the magnetoplasmon energy expressed by the magnetic exciton wavefunction
(see Eq. (\ref{wavefunction-WW})) can be obtained by using
similar notations as above
(let $\vec{n}_\alpha=\vec{n}$ and $\vec{n}_\beta=\vec{n}+\vec{m}$
to denote the hole and electron level indices):
\begin{eqnarray}
\omega_{\vec{m}\vec{n},\rho}(\vec{q}_\perp)
=\Delta E_{\alpha\beta}^0+\Delta E^{\vec{m}\vec{n}}_{bind}(\vec{q}_\perp)
+\Delta E^{\vec{m}\vec{n}}_{RPA}(\vec{q}_\perp)
+\Delta E^{\vec{m}\vec{n}}_{exch}
+\Delta E^{\vec{m}\vec{n}}_{direct}
\label{energy-WW}
\end{eqnarray}
where 
$\Delta E_{\alpha\beta}^0=m_1 \omega_1+m_2 \omega_2-
\omega_z(\sigma_\beta-\sigma_\alpha)$
is the noninteracting energy gap between the two levels, and
\begin{eqnarray}
\Delta E^{\vec{m}\vec{n}_\alpha}_{bind}(\vec{q}_\perp)&=&
-\frac{1}{2\pi l_0^2}\int d\Delta\vec{r}\,
V(\Delta\vec{r}-l_0^2\vec{q}_\perp\times\hat{z})\int
dZ\left|\Psi_{\vec{n}_\beta,\,\vec{n}_\alpha}
(\Delta x,\Delta y,Z,\Delta z)\right|^2
\label{WW-bind}\\
\Delta  E_{RPA}^{\vec{m}\vec{n}_\alpha}(\vec{q}_\perp)&=&\frac{2}
{2\pi l_0^2}\int d\Delta\vec{r}\, V(\Delta\vec{r})\,
e^{i(q_x\Delta x+q_y\Delta y)} \int dZ\,
\Psi_{\vec{n}_\beta,\,\vec{n}_\alpha}(-q_yl_0^2,q_xl_0^2,Z+\Delta z/2,0)
\nonumber \\ &&\times
{\Psi^\ast_{\vec{n}_\beta,\,\vec{n}_\alpha}}(-q_yl_0^2,q_xl_0^2,Z
-\Delta z/2,0)
\label{WW-RPA}\\
\Delta E^{\vec{m}\vec{n}_\alpha}_{exch}
&=&\frac{-1}{2\pi l_0^2}\int d\Delta\vec{r}\,
V(\Delta\vec{r})\int dZ
\left[\Psi_{\vec{n}_\beta,\,\vec{n}_\beta}
(\Delta x,\Delta y,Z,\Delta z)
\sum_{\vec{l}_\beta} {\Psi^\ast_{{\vec{l}_\beta},\, {\vec{l}_\beta}}}
(\Delta x,\Delta y,Z,\Delta z)\right.
\nonumber\\
&&\hspace{1cm}\left.
-\Psi^{\ast}_{\vec{n}_\alpha,\,\vec{n}_\alpha}(\Delta x,\Delta y,Z,\Delta z)
\sum_{\vec{l}_\alpha}{\Psi_{{\vec{l}_\alpha},\, {\vec{l}_\alpha}}}
(\Delta x,\Delta y,Z,\Delta z) \right]
\label{WW-exchange}
\\
\Delta E^{\vec{m}\vec{n}_\alpha}_{direct}&=&
\frac{1}{2\pi l_0^2}
\int d\Delta \vec{r} \, V(\Delta x,\Delta y,\Delta z)
\sum_{\vec{l}}
\int dZ
\Psi_{\vec{l},\vec{l}\,}(0,0,Z-\Delta z/2,0)
\nonumber \\
&&\hspace{1cm}\times
\left[\Psi_{\vec{n}_\beta,\vec{n}_\beta}(0,0,Z+\Delta z/2,0)
-\Psi_{\vec{n}_\alpha,\vec{n}_\alpha}(0,0,Z+\Delta z/2,0)\right]
\label{WW-direct}
,
\end{eqnarray}
where the summation over $\vec{l}$ means the summation over all occupied
levels with quantum number, $(l_1,l_2)$, and summation over
$\vec{l}_{\alpha(\beta)}$ is the summation of all occupied levels with the same
spin as the state $n_{\alpha(\beta)}$.
The interpretation of these equations is similar to the zero width
situation, except for an extra integration over $z$ coordinates.

Note that Eqs. (\ref{WW-bind}), (\ref{WW-RPA}), (\ref{WW-exchange}) and 
(\ref{WW-direct}) can be transformed to the momentum space by using the
$A$ function defined in Eq. (\ref{A_def}):
\begin{eqnarray}
A_{\vec{n}_\beta\vec{n}_\alpha}(\vec{q})&=&
\int dx\int dz\,e^{-iq_xx-iq_zz}
\Phi_{\vec{n}_\beta}(x-q_yl_0^2/2,z)
\Phi_{\vec{n}_\alpha}(x+q_yl_0^2/2,z) \nonumber\\
&=&
\int dz\,e^{-iq_zz}\Psi_{\vec{n}_\beta,\,\vec{n}_\alpha}
(-q_yl_0^2,q_xl_0^2,z,0),
\end{eqnarray}
so that we obtain
\begin{eqnarray}
\Delta E^{\vec{m},\,\vec{n}_\alpha}_{bind}(\vec{q}_\perp)&=&
\frac{-1}{\Omega}\sum_{\vec{p}}\cos((p_yq_x-p_xq_y)l_0^2)
\,V(\vec{p})A_{\vec{n}_\beta\vec{n}_\beta}^\ast (\vec{p})
A_{\vec{n}_\alpha\vec{n}_\alpha}(\vec{p}) \nonumber\\
\Delta E_{RPA}^{\vec{m},\,\vec{n}_\alpha}(\vec{q}_\perp)
&=&
\frac{2}{2\pi l_0^2L_z}\sum_{p_z} V(q_x,q_y,p_z)
|A_{\vec{n}_\beta\vec{n}_\alpha}(q_x,q_y,p_z)|^2  \nonumber\\
\Delta E^{\vec{m},\,\vec{n}_\alpha}_{exch}
&=&\frac{-1}{\Omega}\sum_{\vec{p}}
V(\vec{p})\left[\sum_{\vec{l}_\beta}
\,|A_{\vec{l}\,\vec{n}_\beta}(\vec{p})|^2 - \sum_{\vec{l}_\alpha}
\,|A_{\vec{l}\,\vec{n}_\alpha}(\vec{p})|^2\right]
\nonumber\\
\Delta E^{\vec{m},\,\vec{n}_\alpha}_{direct}&=&
\frac{1}{2\pi l_0^2L_z}\sum_{p_z}\sum_{\vec{l}}V(p_z)A^\ast_{\vec{l}\vec{l}\,}
(p_z)\left[A_{\vec{n}_\beta\vec{n}_\beta}(p_z)-
A_{\vec{n}_\alpha\vec{n}_\alpha}(p_z)\right]
,
\end{eqnarray}
which are identical to the results we have derived before in section 
\ref{pl_disperion} by noting that $A^\ast_{\vec{n}\,\vec{m}\,}(\vec{q})
=A_{\vec{m}\,\vec{n}\,}(-\vec{q})$.
\section{Analytical expression for
$A^{(0),\sigma\sigma'}_{\vec{n}_\alpha\vec{n}_\beta}(\vec{q})$}
\label{A_function}

The explicit formula for the function
$A^{(0),\sigma\sigma'}_{\vec{n}_\alpha\vec{n}_\beta}(\vec{q})$
we use in this paper
can be evaluated by using the known mathematical properties of 
the generalized Laguerre polynomial.
Since it is defined by the noninteracting wavefunctions, which are
not dependent on the spin index explicitly, we can neglect the spin
index totally here and calculate the orbital integer directly
from Eq. (\ref{A_def0}). Using
Eq. (\ref{noninteracting-wavefunctions}) and Eq. (\ref{wf_i}),
we obtain the following results (for convenience,
let $\vec{n}_\alpha=(n_\alpha,n_\alpha')$
and $\vec{n}_\beta=(n_\beta,n_\beta')$):
\begin{eqnarray}
&&A^{(0)}_{\vec{n}_\alpha\vec{n}_\beta}(\vec{q})=
\sqrt{\frac{n_{\alpha\beta,min}!}{n_{\alpha\beta,max}!}\cdot
\frac{n'_{\alpha\beta,min}!}{n'_{\alpha\beta,max}!}}
\nonumber\\
&&\times
\exp\left[-\frac{\cos^2\theta(q_yl_0)^2+
(\cos\theta q_xl_0-\sin\theta q_zl_0)^2\lambda_1^2}{4\lambda_1}\right]
\exp\left[-\frac{\sin^2\theta(q_yl_0)^2+
(\sin\theta q_xl_0+\cos\theta q_zl_0)^2\lambda_2^2}{4\lambda_2}\right]
\nonumber\\
&&\times
\left(\frac{\mp\cos\theta(q_yl_0)-i
(\cos\theta q_xl_0-\sin\theta q_zl_0)\lambda_1}{\sqrt{2\lambda_1}}\right)
^{m_{\alpha\beta}}
\left(\frac{\mp\sin\theta(q_yl_0)-i
(\sin\theta q_xl_0+\cos\theta q_zl_0)\lambda_2}{\sqrt{2\lambda_2}}\right)
^{m'_{\alpha\beta}}
\nonumber\\
&&\times
L_{n_{\alpha\beta,min}}^{m_{\alpha\beta}}\left(\frac{\cos^2\theta(q_yl_0)^2+
(\cos\theta q_xl_0-\sin\theta q_zl_0)^2\lambda_1^2}{2\lambda_1}\right)
L_{n'_{\alpha\beta,min}}^{m'_{\alpha\beta}}\left(\frac{\sin^2\theta(q_yl_0)^2+
(\sin\theta q_xl_0+\cos\theta q_zl_0)^2\lambda_2^2}{2\lambda_2}\right),
\label{A_explicit}
\end{eqnarray}
where $\pm$ is the sign of $n_\alpha^{(,)}-n_\beta^{(,)}$ for each
bracket and $n^{(,)}_{\alpha\beta,min(max)}\equiv
Min(Max)\{n^{(,)}_\alpha,n^{(,)}_\beta\}$, and
$m^{(,)}_{\alpha\beta}\equiv |n^{(,)}_\alpha-n^{(,)}_\beta|$.
$\lambda_{1,2}=(l_{1,2}/l_0)^2$ are dimensionless parameters.
$L_n^m(x)$ is the generalized Laguerre polynomial.

As for a ZWW, we can let $\omega_0\rightarrow\infty$ and obtain
\begin{eqnarray}
A^{2D}_{n_\alpha n_\beta}(\vec{q}_\perp)&=&\sqrt{\frac{n_{\alpha\beta,min}!}
{n_{\alpha\beta,max}!}}\exp\left[-\frac{q^2l_0^2}{4}\right]
\left(\frac{\pm q_yl_0-iq_xl_0}{\sqrt{2}}\right)^m
L_{n_{min}}^m\left(\frac{q^2l_0^2}{2}\right),
\end{eqnarray}
where $q=|\vec{q}_\perp|$, and all notations are the same as
in Eq. (\ref{A_explicit}) above.


\begin{table}
\begin{tabular}{cccc}
{  $\nu$  } & { noninteracting } & { interacting } & { interacting } \\
& & {(unscreened)} & {(screened)} \\ \hline
6 & 19.8 & 11.1 & 12.2 \\
8 & 19.8 & 10.4 & 11.5 \\
\end{tabular}
\caption{Table of the critical values of the parallel magnetic field,
$B_\|^\ast$, where a first order phase transition occurs from an unpolarized
ground state to a polarized one for the parameters of Ref. [7].
}
\label{critical_B_table}
\end{table}
\noindent


\begin{thebibliography} {}
\bibitem{imprefect_parabolic}
As discussed in Sections \ref{noninteracting} and \ref{screening_finite_width},
the in-plane magnetic field effectively enhances the confinement potential
of a parabolic well. Therefore any small deviation of
the bare confinement potential from a perfect parabolic one is relatively
less important when a strong in-plane magnetic field is applied.
Also if needed, nonparabolic effects can be included in our analysis simply
by calculating the matrix element in the corresponding 
noninteracting basis (see
Appendix \ref{HF_equation_append}), which has to be carried 
out numerically for nonparabolic confinement potentials.
\end{thebibliography}
\end{document}